\documentclass[letterpaper, 11pt]{article}
\pdfoutput = 1

\usepackage{shortcuts}

\usepackage[margin = 2.5cm]{geometry}
\usepackage[outline]{contour}

\begin{document}

\thispagestyle{empty}

\begin{flushright}
\texttt{BRX-TH-6660}
\end{flushright}

\begin{center}

\vspace*{5em}

{\LARGE \bf Abelian Bosonization, OPEs,  and \\ the ``String Scale'' of Fermion Fields}

\vspace{1cm}

{\large \DJ or\dj e Radi\v cevi\'c}
\vspace{1em}

{\it Perimeter Institute for Theoretical Physics, Waterloo, Ontario, Canada N2L 2Y5}\\
{\it Martin Fisher School of Physics, Brandeis University, Waltham, MA 02453, USA}\\
\texttt{djordje@brandeis.edu}\\

\vspace{0.08\textheight}
\begin{abstract}
{\normalsize This paper establishes a precise mapping between lattice and continuum operators in theories of (1 + 1)D fermions. To reach the continuum regime of a lattice theory, renormalization group techniques are here supplemented by a new kind of decimation called \emph{smoothing out}. Smoothing out amounts to imposing constraints that make fermion fields vary slowly in position space; in momentum space, this corresponds to introducing boundary conditions below a certain depth of the Dirac sea. This procedure necessitates the introduction of a second small parameter to describe the continuum limit. This length scale, much larger than the lattice spacing but much smaller than the macroscopic system size, controls the derivative expansions of fields, and hence plays the role of the ``string scale'' in quantum field theory. Smoothing out a theory of Dirac fermions on the lattice provides a transparent, fully lattice-based derivation of the operator product expansion, current algebra, and Abelian bosonization rules in the continuum. Nontrivial high-momentum effects are demonstrated in all these derivations. In particular, the radius of the compact scalar dual to the Dirac fermion is found to vary with momentum, smoothly shrinking to zero at a cutoff set by the ``string scale.'' }
\end{abstract}
\end{center}

\newpage

\vspace*{4em}
\tableofcontents
\thispagestyle{empty}

\newpage

\section{Introduction}

This paper is motivated by two basic questions about quantum field theory (QFT):
\begin{enumerate}
  \item Given a continuum QFT, what lattice QFT (if any) describes its full regularization?
  \item Given a lattice QFT, what continuum QFT  (if any) describes its thermodynamic limit?
\end{enumerate}
Answers to these questions are intimately connected to many important research directions in modern field theory. For instance, given a continuum QFT $\mathfrak T$, we may observe the following scenarios:
\vspace*{-1em}
\begin{figure}[h]
\begin{center}
\begin{tikzpicture}[scale = 1]
  \filldraw[fill = green, draw = red, thick] (-0.4, 2) rectangle +(0.2, 0.2);
  \draw (0, 2.2) node[above] {$\mathfrak{L}_1 \approx \mathfrak L_2$};
  \filldraw[fill = green, draw = red, thick] (+0.2, 2) rectangle +(0.2, 0.2);
  \filldraw[fill = blue, draw = red, thick] (0, 0) circle [radius = 0.1];
  \draw (0, -0.1) node[below] {$\mathfrak{T}$};
  \draw[->, thick] (-0.3, 1.8) -- (-0.1, 0.3);
  \draw[->, thick] (0.3, 1.8) -- (0.1, 0.3);

  \draw (0, -1) node {universality};

  \filldraw[fill = green, draw = red, thick] (3.4, 2) rectangle +(0.2, 0.2);
  \draw (4.5, 2.2) node[above] {$\mathfrak{L}_1 \not\approx \mathfrak L_2$};
  \filldraw[fill = green, draw = red, thick] (5.4, 2) rectangle +(0.2, 0.2);
  \filldraw[fill = blue, draw = red, thick] (4.5, 0) circle [radius = 0.1];
  \draw (4.5, -0.1) node[below] {$\mathfrak{T}$};
  \draw[->, thick] (3.5, 1.8) -- (4.4, 0.3);
  \draw[->, thick] (5.5, 1.8) -- (4.6, 0.3);

  \draw (4.5, -1) node {IR duality};

  \filldraw[fill = green, draw = red, thick] (8.9, 2) rectangle +(0.2, 0.2);
  \filldraw[fill = blue, draw = red, thick] (9, 0) circle [radius = 0.1];
  \draw[->, thick] (9, 1.8) -- (9, 0.3);
  \draw (9, -0.1) node[below] {$\mathfrak{T}$};
  \draw (9, 2.3) node[above] {???};

  \draw (9, -1) node {no known UV completion};
\end{tikzpicture}
\end{center}
\end{figure}

\vspace*{-2em}
In the first case, there exist two lattice theories, $\mathfrak L_1$ and $\mathfrak L_2$, that have the same field content but that differ only by the values of irrelevant couplings. Such lattice QFTs at long distances give rise to the same continuum QFT $\mathfrak T$, and they are said to be in the same universality class. Concrete examples are plentiful and can be found in any field theory textbook, e.g.~\cite{Weinberg:1996kr}.

The second situation is more exciting: two lattice QFTs may have starkly different field content but still yield the same continuum QFT. In this case $\mathfrak L_1$ and $\mathfrak L_2$ are said to be infrared-dual. Many examples can be found e.g.~in the recent review \cite{Senthil:2018cru}, where $\mathfrak L_1$ and $\mathfrak L_2$ are called ``strongly dual.'' This paper will be concerned with Abelian bosonization in $(1+1)$D, which is also an IR duality.

The third interesting situation occurs when there are no known lattice realizations of a given continuum theory. The most famous example of such a situation is provided by Einstein gravity, for which no natural set of lattice degrees of freedom is known. Other simple examples are often provided by QFTs with topological terms in the action. More dramatic examples are nonlagrangian conformal theories, with some specific examples originating in refs.~\cite{Witten:1995zh, Garcia-Etxebarria:2015wns}.\footnote{Seiberg duality \cite{Seiberg:1994pq} is an example that combines the second and third scenario: it is an IR duality of two supersymmetric theories.  Latticizing supersymmetric theories is a daunting task (the literature starts with \cite{Dondi:1976tx}, the first hints of trouble are mentioned in \cite{Banks:1982ut}, and a recent overview is given in \cite{Kadoh:2016eju}). In a sense, however, our ignorance about the UV in this case is tamer than in other examples: the field content of the needed lattice theories is more or less known, and the ``only'' challenge is to find the lattice parameters that will ensure supersymmetry at long distances. If this problem were solved, the two lattice theories corresponding to Seiberg-dual pairs of QFTs would be infrared-dual as described in the second diagram above.}

The inverse question involves similar possibilities. Given a fully regularized theory $\mathfrak L$, two interesting scenarios may happen:
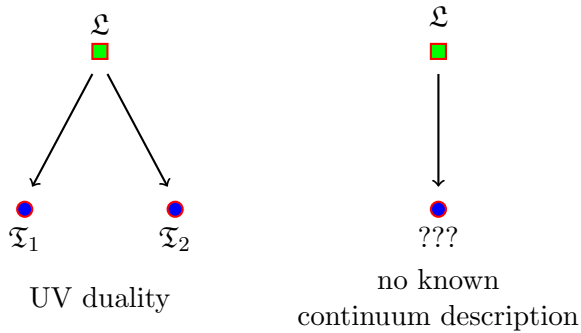
\begin{figure}[h]
\begin{center}
\begin{tikzpicture}[scale = 1]
  \filldraw[fill = green, draw = red, thick] (4.4, 2) rectangle +(0.2, 0.2);
  \draw (4.5, 2.2) node[above] {$\mathfrak{L}$};
  \filldraw[fill = blue, draw = red, thick] (3.5, 0) circle [radius = 0.1];
  \filldraw[fill = blue, draw = red, thick] (5.5, 0) circle [radius = 0.1];
  \draw (3.5, -0.1) node[below] {$\mathfrak{T}_1$};
  \draw (5.5, -0.1) node[below] {$\mathfrak{T}_2$};
  \draw[<-, thick] (3.6, 0.3) -- (4.4, 1.8);
  \draw[<-, thick] (5.4, 0.3) -- (4.6, 1.8);

  \draw (4.5, -1.2) node {UV duality};

  \filldraw[fill = green, draw = red, thick] (8.9, 2) rectangle +(0.2, 0.2);
  \filldraw[fill = blue, draw = red, thick] (9, 0) circle [radius = 0.1];
  \draw[->, thick] (9, 1.8) -- (9, 0.3);
  \draw (9, -0.1) node[below] {$???$};
  \draw (9, 2.3) node[above] {$\mathfrak{L}$};

  \draw (9, -1.2) node (example-align) [align=center] {no known\\ continuum description};
\end{tikzpicture}
\end{center}
\end{figure}

\vspace*{-2em}

In the first situation, there are two distinct continuum theories $\mathfrak T_1$ and $\mathfrak T_2$ that describe the long-distance physics of $\mathfrak L$. Such QFTs are said to be ultraviolet-dual. Typical examples come from pairs of \emph{exactly dual} lattice theories. For instance, an Ising spin system in $(1+1)$D or a Maxwell gauge theory in $(3 + 1)$D are self-dual on the lattice, via Kramers-Wannier duality: their strong- and weak-coupling regimes can be mapped to each other exactly, to all energies, and so the two regimes really describe one and the same lattice theory. The continuum desriptions of these two lattice duality frames give rise to dual QFTs in the infrared. The resulting UV duality is called T-duality in the case of the scalar field in $(1+1)$D, and S-duality (or electric-magnetic duality) for the Maxwell field in $(3+1)$D.

The second situation is very common. Many theories are strongly coupled in the infrared, and we do not know how to describe this macroscopic behavior using  any effective continuum theory. An illustrative example is lattice quantum electrodynamics in $(2 + 1)$D: it is known that when the number of fermion flavors $N_f$ is greater than some critical value $N_f^c$, the theory is described by an interacting conformal theory. At $N_f < N_f^c$ the QFT description is unknown. Worse, even the value of  $N_f^c$ is unknown; see section 4 of \cite{Gukov:2016tnp} for a recent summary of conjectures.

These examples show how deeply the two starting questions penetrate. It should therefore be rather unsettling to learn that the map from a lattice to a continuum QFT has \emph{never} been made completely explicit. Of course, we know that the key role is played by the renormalization group (RG) techniques of Kadanoff and Wilson \cite{Kadanoff:1966wm, Wilson:1971bg}, which highlight the importance of scale invariance in describing continuum limits. Unfortunately, RG is \emph{not enough} to establish a lattice-continuum correspondence. Here is a quick argument. The operators in a lattice theory form a type I operator algebra (in Murray--von Neumann terminology \cite{Murray:1936}), while in a continuum QFT they are expected to form a type III algebra (see e.g.\ \cite{Halvorson2006}). However, no typical RG scheme of a lattice system will change the type of the algebra: it will merely replace a type I algebra with a smaller one.

To give a concrete example, consider the transverse field Ising model in $(1+1)$D. This lattice system is exactly solvable, and at a critical coupling it is described by a nontrivial conformal theory, the Ising CFT. This is perhaps the simplest CFT, and essentially everything is known about it. Despite all this, Wilsonian RG remains silent on how the lattice theory, with an operator algebra generated by Pauli matrices on lattice sites, becomes described by an algebra of conformal operators generated by three primary fields, $\1$, $\sigma$, and $\epsilon$, and their descendants. The lattice-continuum correspondence is only established by closely examining the solution of the lattice theory and manually matching it to a scale-invariant field theory \cite{Kadanoff:1970kz, Koo:1993wz, Milsted:2017csn}.

\textbf{The purpose of the present paper} is to shore up the discussion of the previous two pages and to fill in the gap in the lattice-continuum correspondence by supplementing the hallowed rules of RG with an additional step. This extra step will be called \emph{smoothing out}. If RG is viewed as a projection to an algebra of operators at low spatial momenta --- a procedure that coarse-grains the spatial lattice and returns a smaller lattice with a new, effective Hamiltonian on it --- then smoothing out is a further projection to an algebra of operators \emph{slowly varying} in the coarse-grained position space. This paper will give a precise recipe for smoothing out theories of fermions. The rest of this introduction will give a preview of some general facts about algebras of smooth (slowly varying) operators that will be derived in the main text.

The most important fact is that smooth operators $\O(x)$ are still  operators on a lattice. Let the allowed positions be $x \in \Z$, for simplicity. By construction, all smooth operators will satisfy
\bel{\label{constraint}
  \O(x + 1) = \O(x) + \del \O(x) + \ldots,
}
where $\del \O(x)$ is much smaller than $\O(x)$ in the usual operator norm. The dots indicate even smaller terms. Once all the fields obey such constraints, it is possible to rescale the fields and coordinates, set the lattice spacing to be an infinitesimal quantity, etc. After the rescaling, the smooth operators on the lattice \emph{precisely} behave as the familiar fields of continuum QFT.\footnote{The continuum limit of bosonic lattice theories with continuous target spaces (e.g.~Gaussian scalars or U(1) Maxwell theories) is often defined by restricting attention to field configurations that are close to the ground state of a Hamiltonian, or to the extremum of the action in a Euclidean path integral. This is an energetic way of imposing the smoothness constraint, but it does not provide a satisfactory conceptual picture. For instance, it does not apply to fermionic theories, or to the Ising model.}

The algebra of smooth operators is effectively obtained by imposing the constraint \eqref{constraint} at each point $x$. This means that the smooth algebra is not a direct product of local algebras at each site. The situation is reminiscent of gauge theories, where the Gauss law prevents the factorizability of the gauge-invariant operator algebra (and hence of the gauge-invariant Hilbert space).

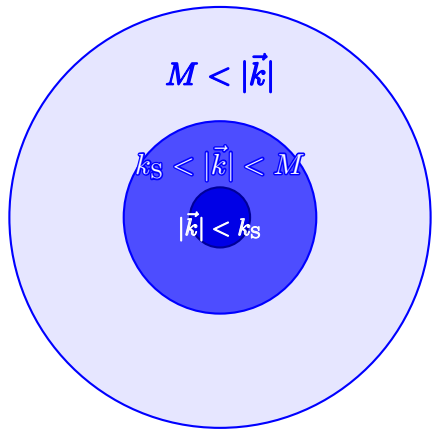
\begin{figure}
\begin{center}
\begin{tikzpicture}[scale = 0.8]

  \filldraw[thick, color = blue, fill = blue!10!white] (0, 0) circle [radius = 3.5];
  \filldraw[thick, blue, fill = blue!70] (0, 0) circle [radius = 1.6];
  \filldraw[thick, blue!60!black, fill = blue!90!black] (0, 0) circle [radius = 0.5];

  \contourlength{0.2pt}
  \draw[white] (0, -0.15) node {\contour{white}{\footnotesize $|\vec k| < k\_S$}};
  \contourlength{0.6pt}
  \draw[white] (0, 0.9) node {\contour{blue}{$k\_S < |\vec k| < M$}};
  \contourlength{0.3pt}
  \draw[blue] (0, 2.4) node {\contour{blue}{\large $M < |\vec k|$}};
\end{tikzpicture}
\end{center}
\caption{\small A momentum space visualization of the two cutoffs, with growing radii corresponding to increasing momentum magnitudes $|\vec k|$. All momentum shells above $M$ are integrated out (Wilsonian RG), while the remaining modes above $k\_S$ are made classical (smoothing).
}
\label{fig momenta}
\end{figure}

Another important fact about the smoothness constraint \eqref{constraint} is that it involves a small parameter that controls the derivative expansion. This parameter is \emph{not} the lattice spacing. In momentum space (with discrete momenta labeled by integers), the lattice spacing is inversely proportional to a momentum cutoff $M \gg 1$; the smoothing out will require the introduction of a smoothness paramater $k\_S$ as a \emph{second} cutoff, at $M \gg k\_S \gg 1$ (see fig.~\ref{fig momenta}). The derivative expansion will then turn out to be controlled by the ratio of cutoffs $k\_S/M$. The new parameter resembles the string scale that controls the derivative expansion in theories of gravity. Thus $k\_S$ can be called the ``string scale,'' the ``smoothness scale,'' or simply the ``second scale.''

Physically, the ``string scale'' will correspond to a scale above which operators stop being quantum. This cutoff cannot be interpreted as integrating out degrees of freedom. (Conversely, the RG scale $M$ is set by integrating out all degrees of freedom at momenta above $M$.) In terms of operators in momentum space, smoothing out corresponds to removing \emph{some} operators at momenta above $k\_S$, ultimately keeping only a set of commuting operators (or classical variables).

This may seem like a bizarre manipulation. However, recall that projecting down to an Abelian algebra on some lattice sites corresponds to imposing specific \emph{boundary conditions} on those sites \cite{Lin:2018bud}. Smoothing out is merely the momentum space analogue. In a ground state of a free fermion theory --- characterized by a ``Dirac sea'' of fermions occupying negative-energy modes --- the smoothing out simply becomes a requirement that the depths of the Dirac sea are to be kept in a fixed configuration. In effect, $k\_S$ is the depth of the accessible, dynamical part of the Dirac sea. This particular method of smoothing out causes individual fermion fields at momenta above $k\_S$ to be projected away, resulting in position-space fermion fields that vary smoothly.

Many nontrivial calculations in continuum QFT require navigating past the smoothness cutoff. The literature almost universally treats it as an unphysical crutch and leaves it implicit or hidden within various calculational techniques, such as point-splitting, normal-ordering, or operator smearing.\footnote{\label{foot KM intro}An example that this paper will analyze in great detail is the derivation of the Kac-Moody current algebra in the theory of free fermions in $(1+1)$D. This calculation, which is also a crucial step in deriving Abelian bosonization, was first done by Schwinger using symmetric point-splitting \cite{Schwinger:1959xd}, and by Lieb and Mattis by (implicitly) imposing the correct boundary conditions deep in the Dirac sea \cite{Mattis:1964wp}. Subsequent versions of the calculation have often relied on various notions of normal-ordering, defined with various degrees of precision. 
When an explicit cutoff \emph{was} used, as in Schwinger's original paper, it often tended to be confused with the lattice spacing, as the catch-all regulator for all UV divergences. However, Luther and Peschel correctly identified the necessary cutoff as the depth of the Fermi sea (or the ``bandwidth'') \cite{Luther:1974}, and Haldane stressed that it ``in no way plays the role of a `cut-off length'\,'' in his seminal paper on Abelian bosonization \cite{Haldane:1981zza}. These authors distinguished the second cutoff from the lattice spacing without revealing its physical role in controlling derivative expansions.} Such techniques are developed anew for each given problem in QFT, and they are usually not motivated by any overarching principle beyond the desire to beat the present calculation into submission. The smoothing procedure given here is designed to be like Wilsonian RG: widely applicable, physically motivated, and overall quite a bit less mysterious than other methods.

A precise definition of smoothing out will be given in \textbf{section \ref{sec smoothing}}. The focus of this paper will be on the theory of a single Dirac fermion in $(1+1)$D, but the smoothing procedure will have a natural generalization to fermions in any number of dimensions. This section will also set the notation for various lattice variables that will be used throughout the paper.

The structure of the algebra of smooth operators will be analyzed in considerable detail in \textbf{section \ref{sec algebra}}. It will be shown that this algebra is \emph{not} a factor, as its center is composed of all the operators above momentum $k\_S$. In other words, this algebra contains many superselection sectors. Each of these sectors (or subfactors) in principle defines a \emph{different continuum limit} of the lattice theory. This paper will focus on sectors naturally corresponding to massless and infinitely massive free Dirac fermions, with the massless (Dirac) sector carefully defined in {\fontseries{b}\selectfont subsection \ref{subsec dirac}}. (The fact that massless and massive regimes correspond to different superselection sectors should be taken as a strong hint that the confined regime of a Yang-Mills theory might not be captured at all by the subfactor that contains information about massless gluons in the continuum.)

Each sector has an algebra characterized by the ``commutator'' between smoothing and operator multiplication. {\fontseries{b}\selectfont Subsection \ref{subsec ope}} will show that these algebraic data, in the Dirac sector and to leading order in $k\_S/M$, become the operator product expansion (OPE) coefficients in a free fermion CFT. This leads to a definition of OPE that does not rely on the gaplessness of the ground state, let alone on the existence of conformal symmetry. The ``string scale'' in this case sets the length scale at which the points in an OPE become so widely separated that the ``singular'' terms become of the same order as the ``regular'' terms that one typically discards in CFT computations.

{\fontseries{b}\selectfont Subsection \ref{subsec KM}} will then turn to the computation of the Kac-Moody structure alluded to in footnote \ref{foot KM intro}. It will be shown that, unlike their lattice progenitors, the smoothed chiral currents of the free fermions have a nontrivial commutation relation in the Dirac state. At low momenta compared to $k\_S$, this will take on the canonical Kac-Moody form, while he result for a general momentum $k$ will be
\bel{\label{KM intro}
  [J^\pm(k), J^\pm(-k')] = \pm \left(k\_S - |k - k\_S|\right)\delta_{k, \, k'}, \quad 0 < k < 2k\_S.
}
This commutator captures the chiral or ABJ anomaly \cite{Adler:1969, Bell:1969}. Another calculation will connect the result \eqref{KM intro} to the fact that the lattice vector and axial currents are not simultaneously on-site \cite{Radicevic:2018zsg}, which is the modern lattice-based point of view on the origin of anomalies \cite{Wen:2013oza}.

As a nontrivial check of the general OPE definition in subsection \ref{subsec ope}, the OPE of chiral currents is computed in {\fontseries{b}\selectfont subsection \ref{subsec JJ ope}}. It is found to match the form expected from CFT Ward identities.

\textbf{Section \ref{sec bosonization}} is devoted to showing that the smoothed algebra of fermions maps to a smoothed algebra of a compact scalar theory. The usual Abelian bosonization \cite{Mattis:1974, Luther:1974,  Coleman:1974bu} is reproduced, which is unsurprising as it is essentially a consequence of the anomalous commutator \eqref{KM intro}. More interestingly, the bosonization rules can be used to infer properties of the smoothed scalar algebra, for which no constructive definition is given in this paper. The dual of a free fermion turns out to be a scalar whose target space size \emph{varies} with momentum as $\sim k\_S^2/k$ at low momenta. The radius of the compact scalar smoothly shrinks to zero at $k = 2k\_S$, reflecting the fact that electron-positron pairs (dual to bosonic excitations) cannot be separated by more than $2k\_S$, the fermion ``bandwidth.''

The paper closes by recasting this new point of view in a more traditional field-theoretic language. The approach presented here is interpreted as a novel way to understand what one means by an operator product in QFT. Some ideas for future work will be presented at the very end.

\textbf{On notation:} With few exceptions, lattice and continuum points are both labeled by $x$, $y$, \ldots \ Lattice operators are \emph{always} written as $\O_x$, with the position (or momentum) in the index. The corresponding smoothed operator is \emph{always} written as $\O(x)$, with the position (or momentum) as an argument. The latter is equal to a continuum operator, up to a rescaling by a constant factor. Finally, \emph{the lattice spacing is set to unity throughout}. This is done to avoid the impression that there are two limits involved, one of the lattice spacing going to zero, another of the number of lattice points going to infinity. The thermodynamic limit in this paper is always taken simply by setting the number of lattice points to be large. This is sufficient for the continuum description to emerge after the smoothing procedure is applied.

\section{Constructing smooth fermion fields} \label{sec smoothing}

Consider a periodic chain of $2N$ sites, with each site hosting a two-dimensional Hilbert space. The algebra of operators acting on this space --- that is, the algebra of all complex $2^{2N}\times 2^{2N}$ matrices --- can be generated by (spinless) complex fermion operators $\psi_v$ and $\psi_v\+$ on all sites $v$. They satisfy the usual anticommutation relations
\bel{\label{ferm alg lat}
  \{\psi_v\+, \psi_u\} = \delta_{vu}, \quad \{\psi_v, \psi_u\} = 0.
}

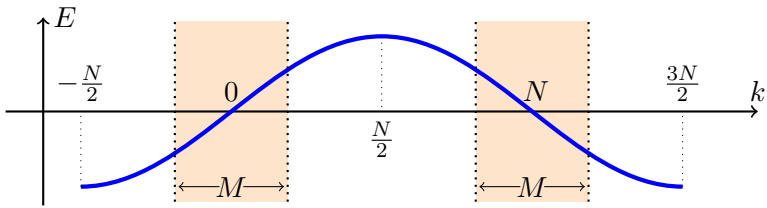
\begin{figure}[h]
\begin{center}
\begin{tikzpicture}[scale = 1]
  \contourlength{1pt}

  \fill[orange!10!yellow!20!pink!50] (1.25, -1.2) rectangle (2.75, 1.2);
  \fill[orange!10!yellow!20!pink!50] (5.25, -1.2) rectangle (6.75, 1.2);

  \draw[->, thick] (-1, 0) -- (9, 0);
  \draw[ultra thick, blue] (0, -1) cos (2,0) sin (4,1) cos (6,0) sin (8, -1);
  \draw (9, 0) node[above] {$k$};

  \draw[->, thick] (-0.5, -1.25) -- (-0.5, 1.25);
  \draw (-0.5, 1.25) node[right] {$E$};

  \draw[dotted, thick] (1.25, -1.2) -- (1.25, 1.2);
  \draw[dotted, thick] (2.75, -1.2) -- (2.75, 1.2);
  \draw[<->] (1.3, -1) -- (2.7, -1) node[midway] {\contour{orange!10!yellow!20!pink!50}{$M$}};

  \draw[dotted, thick] (5.25, -1.2) -- (5.25, 1.2);
  \draw[dotted, thick] (6.75, -1.2) -- (6.75, 1.2);
  \draw[<->] (5.3, -1) -- (6.7, -1) node[midway] {\contour{orange!10!yellow!20!pink!50}{$M$}};

  \draw (2, 0) node[above] {$0$};
  \draw (6.05, 0) node[above] {$N$};

  \draw (0, 0) node[above] {$-\frac N2$};
  \draw[dotted] (0, -1) -- (0, 0);

  \draw (8, 0) node[above] {$\frac {3N}2$};
  \draw[dotted] (8, -1) -- (8, 0);

  \draw (4, 0) node[below] {$\frac N2$};
  \draw[dotted] (4, 1) -- (4, 0);

\end{tikzpicture}
\end{center}
\caption{\small The dispersion of free fermions with Hamiltonian \eqref{UV H k}. The dispersion is periodic ($k + 2N \equiv k$) and approximately linear in the shaded regions of size $M$. Each region corresponds to fermions of a particular chirality. These chiral fermions can be assembled into a Dirac spinor living in a single region of size $M$ \cite{Kogut:1974ag, Susskind:1976jm}.}
\label{fig disp}
\end{figure}

The general form of a free fermion Hamiltonian on this lattice, with a convenient normalization, is
\bel{
  H_0 = \frac12 \sum_{v = 1}^{2N} \left(\e^{\i \alpha} \psi\+_v \psi_{v + 1} + \e^{-\i\alpha} \psi_{v + 1}\+ \psi_v\right).
}
Changing $\alpha$ merely changes the momentum at which the energy is zero, but the physics is the same. For the rest of this paper, $\alpha$ will be set to $\pi/2$. With this choice, introducing the Fourier transform
\bel{
  \psi_v \equiv \frac1{\sqrt{2N}} \sum_{k = -N}^{N - 1} \psi_k \, \e^{\frac{2\pi \i}{2N} v k}
}
gives the Hamiltonian (fig.~\ref{fig disp})
\bel{\label{UV H k}
  H_0 = \sum_{k = -N}^{N - 1} \psi_k\+ \psi_k \sin \frac{\pi k}{N} .
}
A well-known but crucial fact now is that the momentum modes $\psi_k$ obey the same algebraic relations \eqref{ferm alg lat} as the position modes $\psi_v$, and it is possible to associate them with fermion operators acting on a ``dual'' chain of $2N$ sites, each site corresponding to a momentum mode $k$.\footnote{This is \emph{not} the case with, say, Ising spins: while the Pauli matrices $X_v$ and $Z_v$ generate the same algebra as $\psi_v$ and $\psi_v\+$, the momentum space fields $X_k$ and $Z_k$ do not square to the identity or obey the same commutation relations as $X_v$ and $Z_v$ do. This is, ultimately, why this paper is restricted to fermions.}

The momentum modes are all decoupled and this theory is quite trivial on its own. Subtleties arise, however, whenever anything is further done to the system: trying to gauge its symmetries will reveal the chiral anomaly (and the geometric anomaly too \cite{Radicevic:2018zsg}), trying to introduce further couplings will see spinors emerge in degenerate perturbation theory, and trying to build continuum fields out of lattice ones will give rise to anomalous commutators that, in a suitable approximation, form the Kac-Moody algebra. This last point is the topic of the present paper.

What does one mean by a continuum limit of a lattice system?  One part of the story is very well known: if the lattice theory is in a regime where correlation lengths are huge in units of the lattice spacing, then the degrees of freedom on nearby sites are strongly correlated and essentially indistinguishable, and the requisite correlation functions can be obtained using continuum techniques. This is certainly the case in the above theory, where at $N \gg 1$ one finds, for example, $\avg{\psi_v\+ \psi_{v + \Delta v}}_D \sim 1/\Delta v$ for any odd, finite number of lattice spacings $\Delta v$.\footnote{The index $D$ indicates that the vacuum contains a Dirac sea of fermions, i.e.~that only modes at $N < k < 2N$ are occupied. The correlator is actually zero if $\Delta v$ is an even (nonzero) number.} The same power-law arises in a continuum theory of massless Dirac fermions, where $\avg{\Psi\+(x) \Psi(y)} \sim 1/(x - y)$.

This connection between lattice fields $\psi_v$ and continuum fields $\Psi(x)$ is quite vague, however. For example, not all correlation functions can be matched between lattice and continuum. A correlator with $O(N)$ fermion insertions will not be neatly given by a continuum theory. The connection between lattice and continuum \emph{operator algebras} is thus not captured by the above approach of matching correlation functions one-by-one.

The renormalization group (RG) framework, as pioneered by Kadanoff in terms of ``block-spins'' \cite{Kadanoff:1966wm}, may appear to be the sharp connection between lattice and continuum fields. As mentioned in the introduction, this will not be the full story, but it is the right starting point. The idea is to start from the full algebra $\A_N$ generated by the entire set of fermion operators,
\bel{\label{gen AN}
  \left\{\psi_k, \psi_k\+ \right\}_{k = -N}^{N - 1},
}
and to replace it with a smaller one, $\A_M$, generated by
\bel{\label{gen AM}
  \left\{\psi_k, \psi_k\+ \right\}_{k = -\frac M2}^{\frac M2 - 1} \cup \left\{\psi_k, \psi_k\+ \right\}_{k = N - \frac M2}^{N + \frac M2 - 1}, \quad M \ll N.
}
The algebra $\A_M$ only knows about the $M$ modes nearest the ``Weyl points'' $k = 0$ and $k = N$, where the energy is zero (cf.~fig.~\ref{fig disp}). This corresponds to coarsening the lattice from size $2N$ to size $2M$, or to replacing all degrees of freedom within a block of size $N/M$ with a single effective fermion.

The formal manipulations associated to the reduction $\A_N \mapsto \A_M$ are well studied \cite{Lin:2018bud}. Each operator in $\A_N$ can be projected to an operator in $\A_M$. Projecting a density matrix $\rho \in \A_N$ results in a reduced density matrix $\rho' \in \A_M$, which is the unique operator that satisfies
\bel{
  \avg \O \equiv \Tr(\rho \O) = \Tr(\rho' \O) \quad \trm{for\ every\ } \O \in \A_M.
}
In particular, starting from a thermal state $\rho = \frac1Z \e^{-\beta H}$ results in the reduced density matrix $\rho' = \frac1{Z'}\e^{-\beta' H'}$, where $H'$ is the effective Hamiltonian, and $\beta' \equiv \frac MN\beta$ is the rescaled size of the thermal circle that makes sure that space and (Euclidean) time are shrunk by the same factor. The operator $H'$ is \emph{precisely} the effective Hamiltonian that corresponds to the Wilsonian effective action that is so familiar from quantum field theoretic treatments of the renormalization group. In the free theory \eqref{UV H k}, the effective Hamiltonian is (up to normalization) the sum
\bel{
  H'_0 = \sum_{k = - \frac M2}^{\frac M2 - 1} \left(\psi_k\+ \psi_k - \psi_{N + k}\+ \psi_{N + k} \right) \sin \frac{\pi k}N,
}
which is just a truncation of the sum in \eqref{UV H k} to the allowed momenta. Since $|k|\leq \frac M2 \ll N$, this Hamiltonian can also be written as
\bel{\label{IR H k}
  H'_0 = \sum_{k = - \frac M2}^{\frac M2 - 1} \left(\psi_k\+ \psi_k - \psi_{N + k}\+ \psi_{N + k} \right) \frac{\pi k}N + O\left((M/N)^3\right).
}
This is a theory of two species of free fermions with linear dispersions of opposite sign; it is called the free (Tomonaga-)Luttinger model \cite{Tomonaga:1950zz, Luttinger:1963zz}. It is convenient to define
\bel{
  \Psi_k \equiv \bcol{\psi_k}{\psi_{k + N}} \quad \trm{and}\quad \gamma^0 = \bmat0110, \ \gamma^1 = \bmat0{-1}10.
}
The two-component object $\Psi_k$ is a Dirac spinor, with its components $\Psi^\pm_k$ representing fermions of different chiralities. The effective Hamiltonian is, using $\overline\Psi_k \equiv \Psi_k\+ \gamma^0$,
\bel{
  H'_0 = \sum_{k = - \frac M2}^{\frac M2 - 1} \overline\Psi_k \gamma^1 \frac{\pi k}N \Psi_k +  O\left((M/N)^3\right).
}
The effective Hamiltonian is a naive latticization of the continuum Dirac Hamiltonian in one dimension, $-\i \int \overline \Psi \big(\vec\gamma \cdot \vec\del\, \big) \Psi$. This correspondence will be discussed much more below.

From the point of view of position space, the ``infrared'' algebra $\A_M$ is generated by operators
\bel{\label{def Psi x}
  \Psi_x \equiv \frac1{\sqrt{M}} \sum_{k = - \frac M2}^{\frac M2 - 1} \Psi_k \, \e^{\frac{2\pi \i}M k x}, \quad x \in \{1, \ldots, M\}
}
and their conjugates. In terms of the original fermions $\psi_v$, these effective fermions are
\bel{
  \Psi_x = \sum_{v = 1}^{2N} f(v; x) \psi_v,
}
where $f(v; x)$ is a smearing function supported, to a very good approximation, on the interval $v \in \left\{(2x - 1)\frac NM, (2x + 1) \frac NM\right\}$. The components of operators $\Psi_x$ satisfy the same algebraic relations as the $\psi_v$'s. They appear to be natural candidates for continuum fields $\Psi(x)$ mentioned above.

Identifying $\Psi_x$ with $\Psi(x)$ is too fast, however. This is where the current paper builds on the old RG picture of blocking degrees of freedom together. There are many ways to see that additional coarse-graining is needed to reach the continuum. One clue, mentioned in the introduction, is that the reduction $\A_N \mapsto \A_M$ does not change any qualitative features of the algebras, i.e.~they are both type I algebras. A more sophisticated clue is that the ``infrared'' algebra $\A_M$ knows nothing about the most interesting features of continuum field theories, such as current algebras, operator product expansions, or anomalous dimensions. A third clue is that the operators in $\A_M$ do not obey any smoothness conditions, so that there is no sense in which $\Psi_x - \Psi_{x + 1}$ is small.

This last clue also hints towards a construction of a more suitable algebra. The goal is to remove fluctuations below some cutoff length. To this end, consider an algebra $\A_M\^S \subset \A_M$ in which high-momentum modes are not fully integrated out but merely made nondynamical. Concretely, let $1\ll k\_S \ll M$. The \emph{algebra of smooth operators} is generated by
\bel{\label{gen AMS}
  \left\{\psi_k, \psi_k\+, \psi_{k + N}, \psi_{k + N}\+ \right\}_{k = - k\_S}^{k\_S - 1} \cup \left\{\psi_k\+ \psi_k, \psi_{k + N}\+ \psi_{k + N} \right\}_{k = k\_S}^{\frac M2 - 1} \cup \left\{\psi_k\+ \psi_k, \psi_{k + N}\+ \psi_{k + N} \right\}_{k = -\frac M2}^{-k\_S - 1}.
}
In words, all Dirac fermions $\Psi_k$ for $-k\_S \leq k < k\_S$ are dynamical, while the remaining ``high-momentum'' modes only retain their occupation numbers
\bel{\label{def nk}
  n_k \equiv \psi_k\+ \psi_k
}
in the algebra. For future use, it is convenient to define the chiral occupation numbers,
\bel{\label{def nk chiral}
  n^+_k \equiv \psi_k\+ \psi_k = (\Psi^+_k)\+ \Psi^+_k, \quad n^-_k \equiv \psi_{k + N}\+ \psi_{k + N} = (\Psi^-_k)\+ \Psi^-_k.
}

Every operator in $\A_M$ can now be projected to $\A_M\^S$. The projection of a product of operators will be recorded as
\bel{\label{smoothing}
  \O^{(1)}_{x_1} \cdots \O^{(n)}_{x_n} \mapsto \O^{(1)}\cdots \O^{(n)} (x_1, \ldots, x_n).
}
The \emph{key claim} is that the position-space fermions project (up to a rescaling) to the actual continuum fields used in quantum field theory,
\bel{\label{def Psi x smoothed}
  \Psi_x \mapsto \Psi(x) = \frac1{\sqrt M} \sum_{k = -k\_S}^{k\_S - 1} \Psi_k \, \e^{\frac{2\pi \i}M k x}.
}
These fields are smooth, in the sense that they obey the operator equation
\algns{\label{Psi smoothness}
  \Psi(x + 1)
  &= \frac1{\sqrt M} \sum_{k = -k\_S}^{k\_S - 1} \Psi_k \, \e^{\frac{2\pi \i}M k x}\left(1 + \frac{2\pi \i}M k + O\left((k\_S/M)^2\right)\right)\\
  &\equiv \Psi(x) + \hat\del \Psi(x) + O\left((k\_S/M)^2\right).
}
Here $\hat\del$ is a \emph{formal} derivative w.r.t.~$x$, and its action is simply defined as $\hat\del\, \e^{\i \alpha x} \equiv \i \, \alpha\, \e^{\i \alpha x}$.

Why was the introduction of $\A_M\^S$ even needed to get a smoothness relation like \eqref{Psi smoothness}? Starting from $\psi_v \in \A_N$ and projecting it to $\A_M$ would have yielded an operator $\psi(v)$ with an analogous smoothness condition. The point here is that all $\Psi(x)$'s are linearly independent basis vectors of the algebra $\A_M\^S$, and there are $M$ of them, the same number as the number of independent basis vectors $\Psi_x$ of $\A_M$. The RG reduction $\A_N \mapsto \A_M$ does not preserve the number of linearly independent basis vectors in position space, i.e.~not all operators $\psi(v)$ obtained by projecting from $\psi_v$ would have been independent.

Before proceeding to study the detailed algebraic structure of $\A_M\^S$, here are several comments:
\begin{itemize}
  \item Recall that the lattice spacing is set to unity throughout. It is common to introduce a UV spacing $a$ that scales as $1/N$ in the thermodynamic limit $N \rar \infty$, but this crutch is not strictly necessary and it is possible to work directly with $N$.
  \item The standard Wilsonian RG corresponds just to the reduction $\A_N \mapsto \A_M$. This changes the effective lattice spacing (and hence the RG scale)  from $a \sim 1/N$ to $a' \sim 1/M$. The size of the new lattice $M$ is the Wilsonian ``sliding scale.''
  \item An important property of the above construction is that the differentials of fields do not scale as $1/M$ but rather as $k\_S/M$. This ratio controls derivative expansions in the Hamiltonian, and as such it acts as the field-theoretic analog of the string scale in gravity.
  \item The smoothing of operators $\A_M \mapsto \A_M\^S$ is not a homomorphism. This is why the smoothing is recorded in the form \eqref{smoothing}, and one generically has
      \bel{
        \O^{(1)}\cdots \O^{(n)} (x_1, \ldots, x_n) \neq \O^{(1)}(x_1) \cdots \O^{(n)} (x_n).
      }
      For example, at momenta $|k| > k\_S$, the occupation number $n_k^\pm$ remains unchanged after smoothing, even though each of its factors $\Psi^\pm_k$ and $(\Psi_k^\pm)\+$  is individually projected to zero.
  \item The Hamiltonian \eqref{IR H k} of free fermions only depends on the fermion densities $n_k^\pm$. It is thus insensitive to the smoothing out $\A_M \mapsto \A_M\^S$.\footnote{It is the smoothing out that justifies writing the Hamiltonian density as a one-derivative term in position space.} This is not going to be true for more general, interacting Hamiltonians.
  \item This smoothing out should be considered as a novel type of coarse-graining (or RG flow). Within the standard field-theoretic framework in the continuum, smoothing out is not discussed because it is \emph{always} implicitly present when one starts from a continuum QFT and performs, say, momentum-shell RG. In the language of this paper, such an RG procedure means that one starts with a theory with fixed cutoffs $k\_S$ and $N$, and then integrates out high-momentum operators, landing on a theory with cutoffs $k\_S$ and $M < N$.

      As a consequence, every discussion of continuum RG should specify what happens when the RG scale falls below the ``string scale'' $k\_S$ but still remains above  the mass gap. In this situation fields are no longer smoothly varying, and the effective action contains terms with all possible numbers of derivatives simply because the approximation \eqref{Psi smoothness} is no longer valid.
  \item It is not yet obvious how the smoothing out should be defined for bosonic degrees of freedom. The present study of bosonization will shed more light on this.
  \item In terms of Hilbert spaces, there are two consequences of the smoothing out:
      \begin{enumerate}
        \item The pure density matrices that belong to $\A_M\^S$ do \emph{not} have well defined fermion occupation numbers in position space. In other words, no eigenstates of operators $J_x^\pm \equiv (\Psi^\pm_x)\+ \Psi^\pm_x$ are allowed states in the continuum. (These operators --- the chiral currents --- are going to take center stage in the following sections.)
        \item In momentum space, eigenstates of occupation numbers $n_k^\pm$ are present in the smoothed space. However, for $|k|>k\_S$, such eigenstates are \emph{classical}. In the free theory their occupation numbers are constants of motion, and these fixed values can be thought of as \emph{boundary conditions} at the edges of the physical momentum space.
      \end{enumerate}
\end{itemize}

\section{The algebra of smooth operators} \label{sec algebra}

\subsection{The Dirac superselection sector} \label{subsec dirac}

The goal of this section is to analyze the algebra $\A_M\^S$ generated by operators \eqref{gen AMS}. The first step is to note that it possesses a large center: all operators $n_k^\pm$, defined in \eqref{def nk chiral}, belong to the center of $\A_M\^S$ when $-M \leq k < - k\_S$ or $k\_S \leq k < M$. Thus the effective theory of smoothed fermions has a large number of superselection sectors labeled by the eigenvalues of $n_k^\pm$ at high momenta. As pointed out at the end of the previous section, the fermions at momenta above $k\_S$ are not dynamical if the UV theory is free.\footnote{If the UV theory is not free, or more precisely if $[H, n_k^\pm] \neq 0$ for $|k| > k\_S$, then these high-momentum modes will inherit a nontrivial time-dependence in the effective theory, and in particular they may be coupled to each other and to low-momentum modes. This dynamics will \emph{not} be given by unitary evolution by any Hamiltonian from $\A_M\^S$, as this algebra contains no operators that can change the high-momentum occupation numbers.} This means that the free fermion theory can be analyzed sector-by-sector, with each sector labeled by $2(M - 2k\_S)$ constant numbers $n_k^\pm \in \{0, 1\}$.

A particularly natural superselection sector is the one in which precisely the modes with \emph{negative} energy have $n_k = 1$. This is the sector to which the Dirac sea state $\qvec D$ belongs. The Dirac sea is the ground state of the free theory \eqref{UV H k}; it is the eigenstate of all the $n_k$'s, and it satisfies
\bel{
  n_k \qvec D = \left\{
                  \begin{array}{ll}
                    \qvec D, & N < k < 2N; \\
                    0, & \hbox{else.}
                  \end{array}
                \right.
}
This state, of course, remains pure upon coarse-graining from $\A_N$ all the way down to $\A_M\^S$. In terms of the chiral occupation numbers $n_k^\pm$ with $-\frac N2 \leq k < \frac N2$, the Dirac sea condition is
\bel{\label{def Dirac}
  n_k^+ \qvec D = \left\{
                  \begin{array}{ll}
                    \qvec D, & k < 0; \\
                    0, & k \geq 0,
                  \end{array}
                \right.
  \qquad
  n_k^- \qvec D = \left\{
                  \begin{array}{ll}
                    \qvec D, & k > 0; \\
                    0, & k \leq 0.
                  \end{array}
                \right.
}

There is actually a fourfold degeneracy here, as the fermion modes with $k = 0 \equiv 2N$ and $k = N$ both have zero energy even when occupied. Thus there are four ways to define the Dirac sea. The calculations in this section are valid for any of these choices, as all degenerate states belong to the same superselection sector (or to the same subfactor) within $\A_M\^S$. This is the \emph{Dirac superselection sector}.

This discussion is trivial, but also very important. Any different choice of superselection sectors will, in general, lead to a different continuum field theory. In particular, the current algebras of these other field theories will generically have improperly quantized central extensions. 


\subsection{Operator product expansions} \label{subsec ope}

Consider two operators that belong to $\A_M$ but not to $\A_M\^S$. As mentioned above, the product of such operators generically does not commute with the smoothing out. This is how the algebra of smooth operators encodes nontrivial operator product expansions (OPEs).

To see this, consider the smoothing out of the product of two fields $(\Psi_x^\alpha)\+$ and $\Psi_y^\beta$ from eq.~\eqref{def Psi x},
\bel{\label{Psi Psi smoothed}
  (\Psi^\alpha)\+ \Psi^\beta(x, y) = \frac1M \sum_{k,\, l \in \trm{BZ}\_S} (\Psi^\alpha_k)\+ \Psi^\beta_l \, \e^{\frac{2\pi\i}M (ly - kx)} + \frac1M \sum_{k \notin \trm{BZ}\_S} \delta^{\alpha\beta} n^\alpha_k \, \e^{\frac{2\pi\i}M k (y -x)}.
}
In this equation there is no summation over repeated spinor indices like $\alpha$, and $k \in \trm{BZ}\_S$ denotes the restriction of the Brillouin zone $-\frac M2 \leq k < \frac M2$ to momenta $-k\_S \leq k < k\_S$. The sum over $\trm{BZ}\_S$ is precisely $\Psi^\alpha(x)\+ \Psi^\beta(y)$, the product of smoothed fermion fields given in \eqref{def Psi x smoothed}. The second term depends only on the superselection sector and not on the low-momentum dynamics. In other words, this term contains information about the structure of the algebra $\A_M\^S$ that distinguishes it from the type I algebra $\A_M$, and this extra structure can be identified with the OPE data. This presents a robust definition of the general concept of OPEs envisioned by Kadanoff and Wilson \cite{Kadanoff:1969zz, Wilson:1969zs}.

Consider the Dirac sector with $x \neq y$, $\alpha = \beta = +$. The OPE data are contained in the difference
\bel{\label{OPE Dirac}
  (\Psi^+)\+ \Psi^+(x, y) - \Psi^+(x)\+ \Psi^+(y) = \frac1M \sum_{k = -\frac M2}^{-k\_S - 1} \e^{\frac{2\pi\i}M k(y -x)} = \frac{1 - (-1)^{x - y}}{2\pi \i (y - x)} + O\left(\frac{k\_S |x - y|}M \right).
}
This is, in fact, the entire two-point correlation function, and the part that is important at short distances has the same as analytic structure as the known fermion OPE. In the Dirac state, one has
\bel{\label{Psi Psi aux}
  \avg{\Psi^+(x)\+ \Psi^+(y)}_D = \frac{k\_S}M + O\left(\frac{k\_S^2 |x - y|}{M^2} \right),
}
so the sector labels $n_k^\pm$ contain \emph{all} the information about the UV in the smooth limit $k\_S \ll M$.  The remarkable fact is that smoothing out transferred this dynamical information purely into a property of the Dirac superselection sector, and hence into a property of the algebra $\A_M\^S$.

The $O\left(\frac{k\_S |x - y|}M \right)$ terms in the OPE are precisely the nonsingular, uninteresting terms that one discards in conformal field theory (CFT) \cite{Belavin:1984vu, DiFrancesco:1997nk}. They must become important at $|x - y| \gtrsim M/k\_S$.  Thus the ``string scale'' $k\_S$ also sets the \emph{infrared lengthscale} at which CFT techniques break down. In particular, this means that once the ordinary Wilsonian RG decimates the UV regulator $N$ down to $M \sim k\_S$, there is \emph{no} distance $|x - y|$ for which the OPE \eqref{OPE Dirac} works, and the entire edifice of QFT predicated on derivative expansions must come crashing down.

In lieu of further doomsaying, consider what happens in other superselection sectors. A particularly simple one has $n_k^\pm = 0$ for all $k \notin \trm{BZ}\_S$. This sector contains the ground state of a very massive fermion in the UV, as described for example by the simple Hamiltonian
\bel{
  H_{m \rar +\infty} = \sum_{k = -N}^{N - 1} n_k = \sum_{v = 1}^{2N} \psi_v\+ \psi_v = \sum_{k = -\frac N2}^{\frac N2 - 1} \overline \Psi_k \gamma^0 \Psi_k.
}
In the continuum limit (as defined by the reduction to $\A_M\^S$), the OPE of $(\Psi^\alpha)\+$ and $\Psi^\beta$ vanishes.

Note that the OPE is not commutative: the analogous calculation for fields $\Psi^\alpha$ and $(\Psi^\beta)\+$ reveals a nonzero OPE in this sector. It is given by $\delta^{\alpha\beta} \delta_{x, y}$, up to $O\left(\frac{k\_S |x - y|}M \right)$ terms. This same OPE is found for the original operators $(\Psi^\alpha)\+$ and $\Psi^\beta$ in the opposite superselection sector with all $n_k = 1$. This sector is found e.g.~in the ground state of the fermion with a large \emph{negative} mass,
\bel{
  H_{m \rar -\infty} = - \sum_{v = 1}^{2N} \psi_v\+ \psi_v = - H_{m \rar +\infty}.
}

There is an important lesson here: if the superselection sector is always chosen to contain the ground state, then changing the sign of the fermion mass in the UV results in a qualitatively different continuum field theory in terms of its algebra of smooth operators. This difference is pretty benign in this example. However, it is natural to expect it to be a crucial ingredient in the operator-algebraic understanding of, for instance, the ``parity anomaly'' of fermions in $2+1$ dimensions, where different signs of the fermion mass lead to different topological field theories in the IR \cite{Witten:2016cio}.

It is now high time to introduce some special notation for the OPE. Given two operators $\O^{(1)}_{x_1}, \O^{(2)}_{x_2} \in \A_M$, their OPE is defined to be the smooth operator
\bel{\label{OPE}
  \O^{(1)}_{x_1} \times \O^{(2)}_{x_2} \equiv \O^{(1)}\O^{(2)}(x_1, x_2) - \O^{(1)}(x_1) \O^{(2)}(x_2).
}
Note that it is tempting to define the OPE to be a product purely on $\A_M\^S$, but this is impossible at this stage: the smoothed product must know what the original $\A_M$ operators were. The only way to make sense of such a product on $\A_M\^S$ would be to define a canonical uplift $\A_M\^S \mapsto \A_M$, and then to compose it with the OPE \eqref{OPE}. This option will be explored at the end of the paper, when discussing the connections between the present framework and QFT as usually understood.

Keep in mind that this definition of an OPE depends on the choice of the superselection sector. In the free fermion theory studied so far, there are $2^{2(M - 2k\_S)}$ sectors, each defining an OPE in a different continuum theory. Of course, not all these theories will be interesting.

What is a general superselection sector like, in terms of OPEs? For two fermion fields, eq.~\eqref{Psi Psi smoothed} shows that their OPE is always proportional to the identity operator,
\bel{
  (\Psi^\alpha_x)\+ \times \Psi^\beta_y = \frac{\delta^{\alpha\beta}}M \sum_{k \notin \trm{BZ}\_S}n^\alpha_k \, \e^{\frac{2\pi\i}M k (y -x)}.
}
The examples so far can be summarized as follows:
\algns{
  \trm{Dirac\ sector}:& \qquad (\Psi^\alpha_x)\+ \times \Psi^\beta_y = \delta^{\alpha\beta} \frac{1 - (-1)^{x - y}}{2\pi \i (y - x)} + O\left(\frac{k\_S |x - y|}M \right),\\
  \trm{Positive\ mass\ sector}: & \qquad (\Psi^\alpha_x)\+ \times \Psi^\beta_y = 0,\\
  \trm{Negative\ mass\ sector}: & \qquad (\Psi^\alpha_x)\+ \times \Psi^\beta_y = \delta^{\alpha\beta} \delta_{x,\, y} + O\left(\frac{k\_S |x - y|}M \right).
}
These are all instances of the general situation in which the ground state is a Fermi sea $\qvec F$ specified by a momentum $k\_F$, such that
\bel{\label{def F}
  n_k^+ \qvec F = \left\{
                  \begin{array}{ll}
                    \qvec F, & k < k\_F; \\
                    0, & k \geq k\_F,
                  \end{array}
                \right.
  \qquad
  n_k^- \qvec F = \left\{
                  \begin{array}{ll}
                    \qvec F, & k > k\_F; \\
                    0, & k \leq k\_F.
                  \end{array}
                \right.
}
If $|k\_F| < k\_S$, the ground state still lives in the Dirac sector of the smooth algebra. However, if $|k\_F| > k\_S$, then the OPE changes. In particular, dialing $k\_F$ from $\pm k\_S$ to $\pm M/2$ precisely interpolates between the Dirac and massive sectors shown above. For instance, take $k\_F = \lambda M$ for some $\lambda \ll 1$ that is not too small, so that $k\_F \gg k\_S$ holds. Then the Dirac sector OPE \eqref{OPE Dirac} receives a correction $\frac{k\_F - k\_S}M \sim \lambda$ that becomes  important at distances $|x - y| \sim \lambda^{-1}$. In this setup it is now possible to dial $\lambda$ until it becomes $O(1)$ and completely overshadows the $1/(x - y)$ behavior at all scales.

The entire analysis here applies to arbitrary operators, not just to individual fermion fields. All examples so far had OPEs proportional to the identity operator. This is not the case in general. The OPE content of two composite operators will be given by a kind of Wick contraction of generators that constitute the given operators. This is so because the operators, once expanded into momentum-space fermions $\Psi^\alpha_k$, can have pairs of these fermions ``contract'' into classical occupation numbers $n_k^\alpha$ at $|k| > k\_S$ that will survive the smoothing out.

Concretely, let
\bel{
  \O^{\vec \alpha,\,\vec \beta}_{\vec x,\, \vec y} \equiv (\Psi_{x_1}^{\alpha_1})\+ \cdots (\Psi_{x_n}^{\alpha_n})\+ \, \Psi_{y_1}^{\beta_1} \cdots \Psi_{y_m}^{\beta_m}.
}
This operator is the most general ``basis element'' of the algebra $\A_M$. In momentum space, it is given by a sum over many different $(n + m)$-point operators $\O^{\vec \alpha,\,\vec \beta}_{\vec k,\, \vec l}$ weighted by $\frac1{M^{(n + m)/2}} \e^{\frac{2\pi\i}M(\vec l \cdot \vec y - \vec k \cdot \vec x)}$.

An operator like $\O^{\vec \alpha,\,\vec \beta}_{\vec x,\, \vec y}$  is \emph{local} if the various points $x_i$ and $y_j$ are all within $O(1)$ lattice spacings away from each other. For simplicity, assume that none of these ``internal'' coordinates of a local operator are equal. Consider the OPE of two local operators,
\bel{\label{OPE general}
  \O^{\vec \alpha,\,\vec \beta}_{\vec x,\, \vec y} \times \O^{\vec \gamma,\,\vec \delta}_{\vec w,\, \vec z}\, ,
}
which are built out of $n + m$ and $n' + m'$ generators, respectively. To find this OPE, express the ordinary product of these operators in momentum space, getting a sum over all possible terms $\O^{\vec \alpha,\,\vec \beta}_{\vec k,\, \vec l} \O^{\vec \gamma,\,\vec \delta}_{\vec p,\, \vec q}$, summed over all possible momenta $\vec k,\vec l, \vec p, \vec q$. (Each such term is a product over $n + n'$ $\Psi\+_k$'s and $m + m'$ $\Psi_k$'s.) Then, construct all possible ``contractions'' $(\Psi_k^\alpha)\+  \Psi^\delta_q \mapsto \delta^{\alpha \delta} \delta_{k,\,q} n_k^\alpha$ by splitting the momentum sums into sub-sums that have pairs of momenta outside the restricted Brillouin zone $\trm{BZ}\_S$, with all other momenta being inside $\trm{BZ}\_S$. It is precisely such sums that contribute to the OPE \eqref{OPE general}. Performing the sum over the momenta inside $\trm{BZ}\_S$ simply inverts the Fourier transform, giving back the smoothed fermion fields, while doing the sum over the momenta outside $\trm{BZ}\_S$ gives nontrivial functions of the coordinates of contracted fields --- the OPE coefficients $C$. The general form of the OPE thus looks like
\algns{\label{OPE example}
  \O^{\vec \alpha,\,\vec \beta}_{\vec x,\, \vec y} \times \O^{\vec \gamma,\,\vec \delta}_{\vec w,\, \vec z}
  &= C^{\alpha_1, \delta_1}(x_1, z_1) \Big[  \Psi^{\alpha_2}(x_2)\+ \cdots \Psi^{\alpha_n}(x_n)\+ \ \Psi^{\beta_1}(y_1)\cdots \Psi^{\beta_m}(y_m)\ \cdot \\
  & \qquad \qquad \qquad \qquad \cdot \Psi^{\gamma_1}(w_1)\+ \cdots \Psi^{\gamma_{n'}}(w_{n'})\+ \ \Psi^{\delta_2}(z_2)\cdots \Psi^{\delta_{m'}}(z_{m'}) \Big] + \\
  &+ C^{\alpha_1, \delta_2}(x_1, z_2) \Big[  \Psi^{\alpha_2}(x_2)\+ \cdots \Psi^{\alpha_n}(x_n)\+ \ \Psi^{\beta_1}(y_1)\cdots \Psi^{\beta_m}(y_m)\ \cdot \\
  & \qquad \qquad \qquad \qquad \cdot \Psi^{\gamma_1}(w_1)\+ \cdots \Psi^{\gamma_{n'}}(w_{n'})\+ \ \Psi^{\delta_1}(z_1) \Psi^{\delta_3}(z_3) \cdots \Psi^{\delta_{m'}}(z_{m'})\Big] + \ldots
}

The good news is that there are only comparatively few different functions $C$ that need to be computed. The two $C$'s shown in the series above are actually the same function, up to an overall sign due to having to commute one fermion field past a lot of others in order to merge it into a $n_k^\alpha$ operator. In fact, this function $C$ is precisely the same $\sim 1/(x - y)$ function found in \eqref{OPE Dirac}. The locality of operators $\O^{\vec \alpha,\,\vec \beta}_{\vec x,\, \vec y}$ means that there exists a scale that separates the $C$ coefficients arising from contractions that are both ``internal'' (within the same $\O$) and ``external''  (between different $\O$'s). The OPEs of greatest relevance for standard QFT involve only ``external'' contractions, with the functions $C$ depending on at least one $\vec x$ or $\vec y$ coordinate, and at least one $\vec w$ or $\vec z$ coordinate. An explicit example will be computed in subsection \ref{subsec JJ ope}.

There is a huge deal of structure in a sum like \eqref{OPE example}. The details are not relevant for the present paper, but exploring this structure is a worthwhile and tractable task. This situation seems to be the lattice analogue of Weinberg's derivation of OPEs using Feynman diagrams, cf.~\cite{Weinberg:1996kr}, section 20.2.

\subsection{Chiral currents and the Kac-Moody algebra} \label{subsec KM}

In order to move the discussion towards more complicated operators, consider the chiral fermion currents\footnote{Strictly speaking, the $J^\pm$ are just charge densities. However, the spatial component of the continuum current corresponding to the density $J^\pm$ is precisely the density $J^\mp$. Thus it is excusable to sloppily refer to the two-component operator $J^\pm$ as a ``current.'' This way it is also possible to skirt the somewhat subtle question of \emph{defining} spatial components of symmetry currents on the lattice.}
\bel{\label{def J}
  J^\pm_x \equiv (\Psi^\pm_x)\+ \Psi^\pm_x = \frac1M \sum_{k,\, l \in \trm{BZ}} (\Psi^\pm_{l - k})\+ \Psi^\pm_l \, \e^{\frac{2\pi\i}M xk} \equiv \frac1M \sum_{k \in \trm{BZ}} J^\pm_k \, \e^{\frac{2\pi\i}M xk}.
}
Both operators are symmetry currents of the free fermion Hamiltonians \eqref{UV H k} and \eqref{IR H k}. It is important to note that their momentum space versions $J^\pm_k$ are not the same as the momentum occupation numbers $n_k^\pm$. The linear combinations $J^+_x \pm J^-_x$ are typically identified with the vector and axial currents $J_x$ and $J\^A_x$. Both chiral and axial/vector currents are ``on-site'' in the language of Dirac spinors; this will be important during the discussion of anomalies below.

A general Luttinger model is obtained by including bilinears of currents into the Hamiltonian. A special case is the Thirring model \cite{Thirring:1958in}, given up to an overall rescaling by
\bel{
  H_g = H_0 +  \sum_{x, y = 1}^M g_{x, y} J^+_x J^-_y = \sum_{k \in \trm{BZ}} \left( \frac{\pi k}M  (n^+_k - n_k^-) + g_k J_k^+ J_k^- \right).
}
More generally, one can define the interacting Luttinger model\footnote{This form of the Luttinger model Hamiltonian is defined following Haldane \cite{Haldane:1981zza}. In the literature this interacting model is sometimes called the \emph{Luttinger liquid}, but this may be a bit of a misnomer. The original work of Haldane defined the Luttinger liquid as the interacting Luttinger model \eqref{def Lutt Model g h} \emph{plus} higher-derivative terms in the Hamiltonian. (A very clear review of this distinction is given in \cite{Voit:1995}.) The distinction is not crucial because Haldane's point was precisely that including higher-derivative terms does not qualitatively change the Luttinger model picture.

Second, in relativistic QFT, the $J^+ J^-$ term corresponds to $(\overline\Psi \gamma^\mu \Psi)^2$, with the index $\mu \in \{0, 1\}$ contracted using the Minkowski metric. The same term gives $(J^+)^2 + (J^-)^2$ if the Euclidean metric is used to contract $\mu$.}
\bel{\label{def Lutt Model g h}
  H_{g, \, h} = \sum_{k \in \trm{BZ}} \left( \frac{\pi k}M  (n^+_k - n_k^-) + g_k J_k^+ J_k^- + h_k \left[(J_k^+)^2 + (J_k^-)^2\right] \right).
}
This theory is exactly solvable \emph{if} the currents are replaced by their smoothed versions. This will be reviewed in the next section. For now, this is merely motivation for considering the currents $J^\pm_k$.

As explained in the introduction, the chiral currents in the continuum exhibit an anomalous commutation relation in the Dirac sea state. These ``Schwinger terms'' \cite{Schwinger:1959xd} in the current commutators amount to a central extension of the current algebra, giving rise to a ubiquitous Kac-Moody structure that cannot be naively expressed on the lattice. The present goal is to derive these anomalous commutators from the algebra of smooth operators.

The commutator of currents $J^\pm_k$ is zero as an operator equation.\footnote{This is the result that Luttinger obtained in his seminal work \cite{Luttinger:1963zz}; Lieb and Mattis later pointed out that this was wrong in the continuum, where the Dirac sea should have a cutoff with particular boundary conditions at the edges of the sea. This cutoff is precisely the ``string scale'' $k\_S$.} However, this is not the case for their smoothed versions
\bel{\label{def J pm}
  J^\pm(k) = \left\{
               \begin{array}{ll}
                 \sum_{l = -k\_S + k}^{k\_S - 1} (\Psi^\pm_{l - k})\+ \Psi^\pm_l, & 0 < k < 2k\_S; \vspace{1.5ex} \\
                 \sum_{l = -\frac M2}^{\frac M2 - 1} n^\pm_l \equiv N^\pm, & k = 0; \vspace{1.5ex}\\
                 \sum_{l = -k\_S}^{k\_S + k - 1} (\Psi^\pm_{l - k})\+ \Psi^\pm_l, & -2k\_S < k < 0.
               \end{array}
             \right.
}
Note that the momentum that $J^\pm(k)$ depends on is really the \emph{difference} between two momenta in $\trm{BZ}\_S$, and as such it is allowed to take values as large as $\pm(2k\_S - 1)$. Note, also, that $k = 0$ is the only exchange momentum for which the modes above $k\_S$ make an appearance; these ``zero modes'' of currents simply count the total number $N^\pm$ of particles of a given chirality.

Evaluating the commutator of smoothed currents $J^\pm(k)$ is \emph{the} computation that all studies of bosonization perform, although the existence of the second cutoff $k\_S$ is typically elided or mentioned only cryptically. Calculations like this one are most efficiently done by using the identity
\bel{
  [\Psi_k\+ \Psi_l, \Psi_p\+ \Psi_q] = \delta_{lp} \Psi_k \+ \Psi_q - \delta_{kq} \Psi_p\+ \Psi_l.
}
The desired commutators for $k, k' > 0$ are then found to be $[J^\alpha(k), J^\beta(k')] = 0$ and
\bel{
  [J^\alpha(k), J^\beta(-k')] = \delta^{\alpha\beta} \hspace{-3ex} \sum_{l = -k\_S + \max(k, k')}^{k\_S - 1} \hspace{-3ex} (\Psi_{l - k}^\alpha)\+ \Psi^\alpha_{l - k'} - \delta^{\alpha\beta} \hspace{-4ex} \sum_{l = -k\_S + k}^{k\_S - \max(k' - k, 0) - 1} \hspace{-2ex} (\Psi_{l - k + k'}^\alpha)\+ \Psi^\alpha_{l}.
}
The expectation value of the nontrivial commutator in the Dirac sea \eqref{def Dirac} is given by
\bel{\label{KM all momenta}
  \avg{[J^\alpha(k), J^\beta(-k')]}_D = \delta^{\alpha\beta} \delta_{k,\, k'} \!\! \sum_{l = - k\_S + k}^{k\_S - 1} \avg{n^\alpha_{l - k} - n^\alpha_l}_D = 
  \alpha \left(k\_S - |k - k\_S| \right)\, \delta^{\alpha\beta} \delta_{k, \, k'}
}
In particular, for $0 \leq k \leq k\_S$, this commutator takes the conventional Kac-Moody form
\bel{\label{KM low momentum}
  \avg{[J^\pm(k), J^\pm(-k')]}_D = \pm k \, \delta_{k, \, k'}.
}
At higher momenta, $k\_S \leq k \leq 2k\_S - 1$, the relation becomes the much less known
\bel{\label{KM high momentum}
  \avg{[J^\pm(k), J^\pm(-k')]}_D = \pm (2k\_S - k)\, \delta_{k, \, k'}.
}

There are a few remarkable properties of this answer:
\begin{itemize}
  \item The calculation does not depend on the choice of the superselection sector. This anomalous commutator \eqref{KM all momenta} is a purely ``low-momentum'' artifact.
  \item The position space version of the Kac-Moody algebra \eqref{KM low momentum} is often quoted as
      \bel{\label{KM pos space}
        \avg{[J^\pm(x), J^\pm(y)]}_D = \mp \frac \i {2\pi} \delta'(x - y).
      }
      The Fourier transform of \eqref{KM low momentum} gives a very smeared out $\delta'$-function, as the momenta do not run all the way to $\pm \frac M2$. The approximation gets much better after using \eqref{KM all momenta} to reflect that the Kac-Moody structure holds only up to $\pm k\_S$ in momentum space.\footnote{\label{foot KM pos space} The Fourier transform of \eqref{KM all momenta}, with momenta running between $- 2k\_S$ and $2k\_S$, is $\frac{\pm 2\i}{M^2} \sin \frac{2\pi k\_S r}{M}  \big(\sin\frac{\pi k\_S r}{M} \big/ \sin\frac{\pi r}{M}\big)^2$ for $r \equiv y - x$. At $M\gg k\_S \gg 1$, this function can be written as $\mp  \frac{\i} c \del_x \delta(x - y)$, where the constant $c$ depends on the definition of the continuum $\delta'$-function. In general, $c$ scales as some power of $k\_S/M$ and the exact form \eqref{KM pos space} is never obtained at $k\_S \ll M$.}
  \item The Kac-Moody structure holds only up to $\pm k\_S$ in momentum space. The $k\_S$ correction \eqref{KM high momentum} at momentum exchanges $k > k\_S$ does not appear to have been explicitly stated in the literature, even though it logically follows from the established understanding of bosonization. This will be discussed in the next section.
  \item Another way of writing down the anomaly \eqref{KM all momenta} is to use the vector  and axial currents
      \bel{\label{def J vector axial}
        J_x = J^+_x + J^-_x, \quad J\^A_x = J^+_x - J^-_x.
      }
      The anomalous commutator in this case is
      \bel{\label{KM vector axial}
        \avg{\left[J(k), J\^A(-k')\right]}_D = 2\left(k\_S - |k - k\_S|\right)\, \delta_{k, \, k'}, \quad 0 \leq k \leq 2k\_S.
      }
  \item As briefly mentioned below eq.~\eqref{def J}, the currents $J^\pm_x$ are on-site, i.e.~each acts only on the four-dimensional Hilbert space at $x$. These lattice operators commute and do not display any anomalous behavior. Nevertheless, their smoothed versions show an anomaly. Naively, this seems to contradict the principe of anomaly matching \cite{tHooft:1979rat}. The contradiction is avoided because there \emph{does} exist a chiral anomaly in the UV. It arises because the \emph{true} vector and axial currents on the lattice are not simultaneously on-site. (See \cite{Radicevic:2018zsg} for a much more detailed discussion.) Remarkably, two distinct pairs of currents in the UV --- one knowing about the anomaly, one being anomaly-free --- can become approximately equal after smoothing out. This will be made more precise in the rest of this section.
\end{itemize}

To clarify the situation, consider again the free Luttinger model, with momentum-space spinors $\Psi_k \in \A_M$ defined on a lattice with $M$ sites. The mutually ``off-site'' currents are built out of spinless fermions on the lattice with $2M$ sites. In position space, these fermions are defined as\footnote{This may seem like a complicated way to define $\psi_\xi$, however it is easy to verify that they satisfy the same algebraic relations \eqref{ferm alg lat} as the fermions $\psi_v$, but are defined in the $\A_M$ algebra and not in the starting $\A_N$ algebra. In other words, these are the effective spinless fermions one gets after applying Wilsonian RG to reduce $\A_N$ to $\A_M$.}
\bel{
  \psi_{\xi} \equiv \frac1{\sqrt{2M}} \sum_{k \in \trm{BZ}\_S} \left( \Psi_k^+ \, \e^{\frac{2\pi\i}{2M}k \xi} + \Psi^-_k \, \e^{\frac{2\pi\i}{2M} (k + M) \xi} \right), \quad 1\leq \xi \leq 2M.
}
The corresponding smooth fields $\psi(\xi) \in \A_M\^S$ satisfy the pleasant relation
\bel{
  \frac1{\sqrt 2}\big(\psi(2x - 1) \pm \psi(2x)\big) = \Psi^\pm(x).
}
In this sense the Dirac spinors can be understood as simple linear superpositions of spinless fermions on pairs of sites, as proposed by Kogut and Susskind \cite{Kogut:1974ag, Susskind:1976jm}. Of course, this heuristic understanding will not persist to high momenta.

The vector and axial currents of these spinless fermions are
\bel{\label{def j j5}
  j_\xi = \psi_\xi\+ \psi_\xi, \quad j\^A_x = \psi_{2x - 1}\+ \psi_{2x} + \psi_{2x}\+ \psi_{2x - 1}.
}
As already mentioned, these densities are not both on-site: the vector current $j_\xi$ is defined on $2M$ sites $\xi$, and the axial current is defined on $M$ pairs of sites, each pair labeled by $x$ and consisting of original sites $\xi = 2x - 1$ and $\xi = 2x$.

The commutator of the currents \eqref{def j j5} is
\bel{
  \left[j_\xi, j\^A_x\right] = \left(\psi_{2x - 1}\+ \psi_{2x} - \psi_{2x}\+ \psi_{2x - 1}\right) \left(\delta_{\xi, 2x - 1} - \delta_{\xi, 2x}\right),
}
and in the Dirac state it is easily evaluated to be
\bel{\label{KM lattice}
  \avg{\left[j_\xi, j\^A_x\right]}_D = -\frac{2\i}\pi \left(\delta_{\xi, 2x - 1} - \delta_{\xi, 2x}\right) + O\left(\frac1M\right).
}
The formal similarity with the continuum correlator \eqref{KM pos space} is now apparent. (In general, the appearance of derivatives of $\delta$-functions in a current algebra immediately signals that the currents are not mutually on-site.)   Note that the prefactor $2\i/ \pi$ arises purely from the UV correlations of two fermions in the Dirac sea; the calculation is closely related to the OPE expansion \eqref{OPE Dirac}. This result can be compared to the position space version of \eqref{KM vector axial}, $\avg{[J(x), J\^A(y)]}_D = - \frac{2\i}{c} \delta'(x - y)$.

Now comes the key fact: the anomalous commutator \eqref{KM lattice} is \emph{also} detected by the smoothed operators $j(\xi)$ and $j\^A(x)$. This is in contrast to the product of two fermion fields \eqref{OPE Dirac}, where the smoothed fermions essentially detected nothing about the two-point function of UV fermions.

This can be demonstrated by a direct computation. The first step is to express the lattice currents in momentum space:
\algns{
  j_\xi
  &= \frac1{2M} \sum_{k,\, l  \in \trm{BZ}}\left[ (\Psi_{l - k}^+)\+ \Psi_l^+ + (\Psi_{l - k}^-)\+ \Psi_l^- + (-1)^\xi (\Psi_{l - k}^+)\+ \Psi_l^- + (-1)^\xi (\Psi_{l - k}^-)\+ \Psi_l^+ \right] \e^{\frac{2\pi\i}{2M}\xi k},\\
  j\^A_x
  &= \frac1{M} \sum_{k,\, l  \in \trm{BZ}} \left[ (\Psi_{l - k}^+)\+ \Psi_l^+ - (\Psi_{l - k}^-)\+ \Psi_l^- + (\Psi_{l - k}^+)\+ \Psi_l^- - (\Psi_{l - k}^-)\+ \Psi_l^+ \right] \e^{\frac{2\pi\i}{M} x k + \frac{2\pi\i}{2M}(l - k)} + \trm{H.c.}
}
Now it is easy to perform the smoothing out, in analogy with \eqref{def J pm}. In terms in which the exchange momentum is positive, $0 < k < 2k\_S$, the sum over $l$ is restricted to $\sum_{l = -k\_S + k}^{k\_S - 1}$ after the smoothing. In terms with $-2k\_S < k < 0$, this sum becomes restricted to $\sum_{l = -k\_S}^{k\_S + k - 1}$.

Notice that the first two terms in these currents contain precisely the same operators that figure in the chiral currents $J^\pm(k)$, eq.~\eqref{def J pm}. The only difference are the complex weights, but in the smoothed regime these difference are negligible. In other words, the following expansions are allowed:
\bel{
  \e^{\frac{2\pi\i}{2M} \xi k} = \e^{\frac{2\pi\i}M \lceil \frac\xi2 \rceil k}  + O\left(\frac{k\_S}M\right), \quad \e^{\frac{2\pi\i}{2M} (l - k)} = 1 + O\left(\frac{k\_S}M\right).
}
For convenience, let $y \equiv \lceil \frac\xi2 \rceil$, so that $\xi = 2y - 1$ and $\xi = 2y$ both give $y$ in the exponent after applying this approximation.

Up to $O(k\_S/M)$ terms, the commutator $\left[j(\xi), j\^A(x)\right]$ contains \emph{exactly the same} anomalous term exhibited by the vector/axial currents in eq.~\eqref{KM vector axial},
\bel{
  \left[j(\xi), j\^A(x)\right] \supset \frac1{M^2} \sum_{k,\, k' \in \trm{BZ}\_S} \left[J(k), J\^A(k') \right] \, \e^{\frac{2\pi\i}{M} (k y + k' x)} + O\left(\frac{k\_S}M\right).
}
However, $\left[j(\xi), j\^A(x)\right]$ has four more terms that are individually nonzero in the Dirac sea. They can be arranged as
\algns{\label{KM aux}
  \left[j(\xi), j\^A(x)\right] \supset \frac{(-1)^\xi}{2M^2} \sum_{\substack{k, \, l  \in \trm{BZ}\_S\\ k', \, l' \in \trm{BZ}\_S}} &\Bigg(
  \left[(\Psi_{l - k}^-)\+ \Psi_l^+, (\Psi_{l' - k'}^+)\+ \Psi_{l'}^- \right] - \left[(\Psi_{l - k}^+)\+ \Psi_l^-, (\Psi_{l' - k'}^-)\+ \Psi_{l'}^+ \right] \Bigg) \times \\
  &\quad \times  \e^{\frac{2\pi\i}{2M} \xi k} \left(\e^{\frac{2\pi\i}M xk' + \frac{2\pi\i}{2M}(l' - k')} - \e^{-\frac{2\pi\i}M xk' - \frac{2\pi\i}{2M}(l' - k')}\right).
}
As before, it is possible to expand  $\e^{\frac{2\pi\i}{2M}(l' - k')} = 1 + O(k\_S/M)$, eliminating the $l$ or $l'$ dependence from the phase factors above. Each commutator in this expression gives fermion bilinears of the form
\bel{
  \delta_{l,\, l' - k'} (\Psi_{l - k}^\alpha)\+ \Psi^\alpha_{l'} - \delta_{l',\, l - k} (\Psi_{l' - k'}^\beta)\+ \Psi^\beta_{l},
}
with $(\alpha, \beta) = (-, +)$ for the first commutator, and $(+, -)$ for the second one. In the Dirac sea, the difference of commutators in \eqref{KM aux} is
\bel{
  \delta_{l,\, l' - k'}\, \delta_{l',\, l - k}\, \avg{n^-_{l - k} + n^-_l - n^+_{l - k} - n^+_l}_D.
}
Summing this function over $l$ gives zero for any $k$. This means that the commutator $\left[j(\xi), j\^A(x)\right]$ contains \emph{precisely} the same information as the commutator of chiral currents \eqref{KM all momenta} or its vector/axial version \eqref{KM vector axial}.

As advertised, the smoothed currents $j(\xi)$, $j\^A(x)$ turn out to have nontrivial correlations, unlike the smoothed fermion fields $\Psi\+(x)$, $\Psi(y)$ in \eqref{Psi Psi aux}. Moreover, notice how the distinction between coordinates $\xi = 2y$ and $\xi = 2y - 1$ disappeared from the commutator after smoothing. This calculation gives a rigorous explanation of how spinless fermions on the lattice with $2M$ sites can behave as if they lived on the lattice with $M$ sites, with the off-site symmetries $j_\xi$, $j\^A_x$ now appearing as perfectly on-site (yet anomalous) symmetries $j(y)$, $j\^A(x)$. This also explains how the smoothing of nonanomalous currents $J_x^\pm$ can uncover an anomaly: since the smoothing out is a many-to-one map, in this case the smoothing of $J_y$ and $J\^A_x$ happens to coincide with the smoothing of $j_\xi$ and $j\^A_x$ for the purposes of computing the commutators. It would be nice to develop a set of rules that determine when smoothing out a nonanomalous field may get ``contaminated'' by an anomaly.

In general, if two smoothed fields $\O(x)$ and $\~\O(y)$ have a nontrivial correlation function (represented by a dependence on $|x - y|$ at leading order in the smooth limit), this suggests the presence of some kind of nontrivial long-range order, or of a nontrivial RG invariant. In the present case the correlator of smoothed currents captures the chiral anomaly. A similar calculation for smoothed fermion currents in $(2+1)$D should capture the contact terms that arise in the presence of topological gauge fields \cite{Closset:2012vp}.

Perhaps the correlation functions of smooth operators can be useful even in simpler setups. For example, $\avg{\Psi(x)\+\Psi(y)}_D$, given in \eqref{Psi Psi aux}, does not depend on $|x - y|$ at leading order. However, the coefficient of the leading term of this correlator continuously changes from zero (positive mass) to $2k\_S/M$ (negative mass) as the mass is dialed. It would be interesting if this coefficient could be given a physical interpretation.

\subsection{The current-current OPE} \label{subsec JJ ope}

A nontrivial sanity check of the philosophy so far is the computation of the current-current OPE, for which CFT techniques based on Ward identities give the generic form \cite{DiFrancesco:1997nk}
\bel{\label{OPE CFT}
  J^\alpha(x)  J^\beta(y) \sim \frac{\delta^{\alpha\beta}}{(x - y)^2}  + \trm{regular\ terms}.
}
To derive this by smoothing out the lattice theory, follow the prescription from the end of subsection \ref{subsec ope}, around eq.~\eqref{OPE example}. Start from the definition \eqref{OPE}, applied to two chiral currents \eqref{def J}:
\bel{
  J_x^\alpha \times J_y^\beta = J^\alpha J^\beta(x, y) - J^\alpha(x) J^\beta(y).
}
The Fourier expansions of the two currents will be denoted as
\bel{
  J^\alpha_x = \frac1M \sum_{k,\, l \in \trm{BZ}} (\Psi^\alpha_k)\+ \Psi^\alpha_l\, \e^{\frac{2\pi\i}M (l - k) x}, \quad J^\beta_y = \frac1M \sum_{p,\, q \in \trm{BZ}} (\Psi^\beta_p)\+ \Psi^\beta_q \, \e^{\frac{2\pi\i}M (q - p) y}.
}
The product of lattice operators, $J_x^\alpha J_y^\beta$, is a sum over all possible terms of form $(\Psi^\alpha_k)\+ \Psi^\alpha_l(\Psi^\beta_p)\+ \Psi^\beta_q$. The terms that appear in the OPE are those in which there is at least one pair of momenta that are equal and lie outside the smoothed Brillouin zone BZ$\_S$. Thus it is necessary to study all possible ``Wick contractions'' of fermion fields into occupation numbers $n_k^\pm$ that can survive the unhospitable environment outside BZ$\_S$.

There are two qualitatively different kinds of contractions. First, there are the ``internal'' ones, contracting fermion fields from the same current operator. These give, in the Dirac sector,
\algns{
  J_x^\alpha \times J_y^\beta
  &\supset \frac1{M} \sum_{k \notin \trm{BZ}\_S} n^\alpha_k J^\beta(y) + \frac1{M} \sum_{p \notin \trm{BZ}\_S} n^\beta_p J^\alpha(x) + \frac1{M^2} \sum_{k,\, p \notin \trm{BZ}\_S} n^\alpha_k n^\beta_p \\
  &= \frac12 \left(J^\alpha(x) + J^\beta(y)\right) + \frac14 + O\left(\frac{k\_S}M \right).
}
These are all ``regular'' terms, with no interesting dependence on the position.

More interesting are the ``external'' contractions. The terms they engender within the OPE are
\algns{
  J_x^\alpha \times J_y^\beta
  &\supset - \frac{\delta^{\alpha\beta}}{M^2}\!\! \sum_{\substack{k \notin \trm{BZ}\_S\\ p,\, q \in \trm{BZ}\_S}} \!\! n_k^\alpha\, (\Psi_p^\alpha)\+ \Psi_q^\alpha \left( \e^{\frac{2\pi\i}M [(q - k)x + (k - p)y]} +  \e^{\frac{2\pi\i}M [(q - k)y + (k - p)x]} \right) \\
  &\qquad \qquad \qquad - \frac{\delta^{\alpha\beta}}{M^2}\!\! \sum_{k,\, p \notin \trm{BZ}\_S} n_k^\alpha n_p^\alpha \e^{\frac{2\pi\i}M (k - p) (y - x)} + \frac{\delta^{\alpha\beta}}{M^2}\!\! \sum_{k,\, p \notin \trm{BZ}\_S} n_k^\alpha \e^{\frac{2\pi\i}M (k - p) (y - x)}.
}
A few things are worth noting. The minus signs in front come from having to anticommute $(\Psi_p^\beta)\+$ past $\Psi_l^\alpha$ to achieve the contraction. The same anticommutation gives rise to the last sum in the expression. This particular sum is the boring one, as it evaluates to $\frac{k\_S}M + O\left(\frac{k\_S |x - y|}M \right)$.

The other two sums are much more interesting. The second one is
\bel{
  -\frac{\delta^{\alpha\beta}}{M^2}\!\! \sum_{k,\, p \notin \trm{BZ}\_S} n_k^\alpha n_p^\alpha \e^{\frac{2\pi\i}M (k - p) (y - x)} = -\delta^{\alpha\beta} \frac{(1 - (-1)^{x - y})^2}{4\pi^2 (y - x)^2} + O\left(\frac{k\_S |x - y|}M \right).
}
The crucial part is the scaling as $\sim 1/(y - x)^2$: this gives the most singular term in the CFT result \eqref{OPE CFT}. The constants can be changed by rescaling currents, and the usual oscillation in the numerator is lost upon integrating against any smooth test function.

The most interesting is the first sum. It actually contains two terms, and one of them is
\bel{\label{aux term}
  -\frac{\delta^{\alpha\beta}}{M^2} \sum_{k \notin \trm{BZ}\_S} n_k^\alpha \, \e^{\frac{2\pi\i}M (x - y) k} \sum_{p,\, q \in \trm{BZ}\_S} (\Psi_p^\alpha)\+ \Psi_q^\alpha \, \e^{\frac{2\pi\i}M [(q - p)x - q (x - y)]}.
}
The sum over $k$ is familiar and results in a $\sim 1/(x - y)$ function. The sum over $p$ and $q$ can be evaluated by expanding in $q|x - y|/M$, using the fact that $|q| < k\_S$. The leading term in $\sum_{p, q}$ simply becomes $J^\alpha(x)$, so the entire expression \eqref{aux term} is $\sim J(x)/(x - y)$ to leading order. The subleading term in $\sum_{p, q}$ can be written as $\sim (x - y) \hat\del_x J^\alpha(x)$; the prefactor cancels against the $1/(x - y)$ coming from the sum over $k$. Thus this subleading term is a regular term that scales as $k\_S/M$.

The other term in the first sum is exactly equal to this one, but has the opposite sign. Thus the first sum, in its entirety, actually \emph{vanishes} up to $k\_S/M$ effects. The appropriate $J(x)/(x - y)$ term will only arise in the presence of multiple fermion flavors. This will not be analyzed in this work, but the above calculation clearly shows how this will happen.

To conclude, the OPE of two chiral currents is found to be
\bel{\label{OPE JJ}
  J_x^\alpha \times J_y^\beta = -\delta^{\alpha\beta} \frac{(1 - (-1)^{x - y})^2}{4\pi^2 (y - x)^2} + \frac12 \left(J^\alpha(x) + J^\beta(y)\right) + \frac14 + O\left(\frac{k\_S|x - y|}M \right).
}
This is, indeed, the expected scaling of the leading OPE term \eqref{OPE CFT}. Note that this approach allows the complete determination of regular terms, should they ever be of interest. These regular terms need not all be subleading: for instance, the $1/4$ becomes comparable to the singular piece $1/(x - y)^2$ when $x - y$ is a small integer.

\newpage

\section{Abelian bosonization at finite ``string scale''} \label{sec bosonization}

\subsection{Compact scalars}

Abelian bosonization associates the smoothed conserved currents of a fermion theory to smoothed operators in a compact scalar theory. Recall that the compact scalar on a lattice can be obtained as the $K \rar \infty$ limit of the $\Z_K$ clock model. (The $\Z_2$ clock model is simply the Ising model.) The operators that generate the clock algebra are generalizations of Pauli matrices $Z_v$ and $X_v$ on each site $v$. All operators on different sites commute, and on the same site these operators satisfy
\bel{
  Z_v X_v = \omega X_v Z_v, \quad X_v\qvec{\omega^n}_v = \omega^n \qvec{\omega^n}_v, \quad Z_v\qvec{\omega^n}_v = \qvec{\omega^{n-1}}_v
}
for $\omega \equiv \e^{\frac{2\pi\i}K}$; the Hilbert space on each site is $K$-dimensional.

The compact scalar algebra is now generated by operators $\phi_v$ and $\pi_v$ on each site, with
\bel{
  X_v \equiv \e^{\i\phi_v}, \quad Z_v  \equiv \e^{\i \pi_v \d\phi} \qquad (\d\phi\equiv 2\pi/K).
}
In states whose wave functions vary smoothly in the target space, the shift operator is $\pi_v \approx -\i \pder{}{\phi_v}$. This ``target space smoothness'' is the crucial addition to the philosophy of smooth algebras that must be taken into account in bosonic theories.

With these conventions, the Ising model
\bel{
  H\_{Ising} = \sum_{v = 1}^N \left(X_v X_{v + 1} + h Z_v\right) \qquad (K = 2)
}
\bel{
  H_K = \sum_{v = 1}^N \left(X_v\+ X_{v + 1} + X_{v + 1}\+ X_v + h(Z_v + Z_v\+)\right) = 2 \sum_{v = 1}^N \left(\cos(\phi_v - \phi_{v + 1}) + h \cos(\pi_v)\right).
}
Upon restricted to smoothly varying fields in both real space and target space, the effective theory is the familiar free scalar
\bel{\label{def H K}
  H'_K \approx \sum_{x = 1}^M \left(\left(\del\phi(x)\right)^2 + h\, \pi^2(x)\right).
}
Theories of this form --- potentially with interacting Hamiltonians, e.g.~with sine-Gordon potentials $\cos\phi(x)$ --- are mapped to continuum fermions by the bosonization rules established in \cite{Luther:1974, Mattis:1974, Coleman:1974bu}.

\subsection{Bosonization of operators}

Instead of specifying a smoothing out of the bosonic algebra that leads to the effective Hamiltonian $H'_K$ in \eqref{def H K}, here the starting point will be the bosonization duality. The simplest way to introduce it is \emph{in medias res}, by positing a map between scalar operators and vector/axial currents \eqref{def J vector axial}:\footnote{Fields $\pi(x)$ will always be written with their arguments, to set them apart from the constant $\pi = \left(\int \d t \, \e^{-t^2}\right)^2$.}
\algns{\label{duality basic}
  \del \phi(x) &= c\_F \left(J(x) - \frac{N\^F}{M}\right),\\
  \pi(x) &= c\_A \left(J\^A(x) - \frac{N\^A}{M}\right).
}
Here the fermion and axial numbers are defined in analogy with the chiral numbers $N^\pm$ from eq.~\eqref{def J pm},
\bel{
  N\^F \equiv J(k = 0) = N^+ + N^-, \quad N\^A \equiv J\^A(k = 0) = N^+ - N^-.
}
The duality \eqref{duality basic} maps the density fluctuations of vector and axial charges to ``elementary'' scalar fields, up to a choice of proportionality constants $c\_F$ and $c\_A$. The anomalous commutation relation, at least in the Dirac sea, now becomes consistent with the canonical commutation of the scalar theory,
\bel{
  \avg{[\del_x\phi(x), \pi(y)]}_D = c\_F\, c\_A \avg{\left[J(x), J\^A(y)\right]}_D = - \i\, \frac{c\_F\, c\_A}{c} \del_x \delta(x - y).
}

The constants $c\_F$ and $c\_A$ can now be chosen to cancel out $c$, the debris due to finite-$k\_S$ effects (see footnote \ref{foot KM pos space}). Note that this prefactor does not appear in the literature, which seems to uniformly assume that the naive position space commutator \eqref{KM pos space} is correct. As forcefully argued in this paper, this assumption cannot be correct if the fields are smooth or if the derivative expansion makes sense;  a ``string scale'' is needed to even \emph{define} such smoothness, and its introduction changes the various anomalous commutators in nontrivial ways.

The physical interpretation of the scalar field is best understood in momentum space. The Fourier transform is easier to do when working with chiral currents, so let
\bel{
  \del \phi_\pm(x) \equiv \frac12 \left(\frac1{c\_F} \del \phi(x) \pm \frac1{c\_A} \pi(x)\right).
}
With this judicious choice of coefficients, bosonization maps these chiral bosonic operators to chiral density fluctuations,
\bel{\label{duality basic pm}
  \del \phi_\pm(x) = J^\pm(x) - \frac{N^\pm}{M}.
}

The r.h.s.~of this map can be expanded into Fourier modes following eqs.~\eqref{def J} and \eqref{def J pm}, giving
\algns{
  J^\pm(x) - \frac1M N^\pm
  &= \frac1M \sum_{k = -2k\_S + 1}^{2k\_S - 1}\!\! J^\pm(k) \e^{\frac{2\pi\i}M kx} - \frac1{M}N^\pm \\
  &=  \hat\del_x \sum_{k = 1}^{2k\_S - 1} \left(\frac{J^\pm(k) }{2\pi \i k} \e^{\frac{2\pi\i}M kx} - \frac{J^\pm(-k) }{2\pi\i k}  \e^{-\frac{2\pi\i}M kx}  \right).
}
Recall that $\hat \del_x$ denotes the formal derivative w.r.t.~$x$ introduced in eq.~\eqref{Psi smoothness}. The scalar $\phi_\pm(x)$ must obey the same smoothness condition, $\hat \del_x \phi_\pm(x) = \del_x \phi_\pm(x)$, for the duality \eqref{duality basic} to be consistent at all $x = 1, \ldots, M$. Thus the Fourier transform of $\phi_\pm$ can be defined as
\bel{
  \phi_\pm(x) \equiv \sum_{k = 1}^{2k\_S - 1} \frac{\sqrt{k_\star}}{2\pi k} \left(b_\pm(k)
  \e^{\frac{2\pi\i}{M} k x} + b_\pm\+(k) \e^{-\frac{2\pi\i}{M} k x} \right)
}
with
\bel{\label{def b}
  b_\pm(k) \equiv \frac{-\i}{\sqrt{k_\star}}J^\pm(k), \qquad k_\star \equiv k\_S - |k - k\_S|.
}
The awkward momentum function $\sqrt{k_\star}$ will make sure that the bosonic fields $b_\pm(k)$ at \emph{all} momenta satisfy the canonical commutation relation for ladder operators in the Dirac sea,
\bel{\label{KM boson}
  \avg{\left[b_\pm(k), b_\pm\+(k')\right]}_D = \pm \delta_{k, \, k'}.
}

It is now clear that $b_+(k)$ and $b_-\+(k)$ act as lowering operators at $0 < k < 2k\_S$. The space of bosonic states in $\A_M\^S$ can thus be imagined as a set of $2k\_S - 2$ harmonic oscillators, one for each momentum $k \neq 0$ of magnitude $|k| < 2k\_S$. By construction, there are no scalar degrees of freedom associated to the $k = 0$ mode of the fermion currents. This means that the duality \eqref{duality basic} maps each superselection sector of the chiral fermion number symmetry (generated by $N^\pm$) to a \emph{different} bosonic theory.\footnote{These are superselection sectors based on the total fermion numbers $N^\pm = \sum_{k = -M/2}^{M/2-1} n^\pm_k$. Even after the superselection sector within $\A_M\^S$ is fixed, the remaining Hilbert space will still split into yet smaller sectors labeled by the total $N^\pm$ values.} The individual fermion operators $\Psi^\pm(x)$ do not map to anything. The bosonic side is in a similar situation. The fields $\phi(x)$ do not have duals under \eqref{duality basic} --- only their derivatives dualize. Accordingly, each sector of the shift symmetry $\phi(x) \mapsto \phi(x) + \lambda$ dualizes separately.

The Hilbert space on which the $b$ fields act cannot support infinitely many excitations. The fermion theory is manifestly finite, with $2^{4k\_S}$ linearly independent dynamical states. This means that there are additional constraints that the ladder algebra \eqref{KM boson} is subject to. These constraints are, by and large, ignored in the literature, as they become relevant at high momenta only.

To get a feel for the kinds of constraints that appear on the bosonic side, consider the action of $b_+(k)$ on the fermionic side. (This is illustrated in great detail in \cite{vonDelft:1998} for low-momentum excitations.) This operator, defined in \eqref{def b}, acts as
\bel{
  b_+(k) \propto \sum_{l = -k\_S + k}^{k\_S - 1} (\Psi^+_{l - k})\+ \Psi^+_l.
}
Starting from a state with definite occupation numbers $n_k^+$, $b_+(k)$ creates a superposition of all possible ``electron-positron'' pairs in which the electron has momentum exactly $k$ less than the positron. In the Dirac sea, all fermion modes below $l = 0$ are occupied by an electron; if a hole in the sea (i.e.~a positron) is created at some momentum $l < 0$, there will be no room at any momentum $l - k < l$ for the corresponding electron, and so $b_+(k)$ annihilates the Dirac sea. However, acting on a highly excited state with lots of electrons at $l > 0$, it will take a lot of applications of $b_+(k)$ until it returns zero.

Conversely, the raising operator $b\+_+(k)$ creates electron-positron pairs with the electron having the greater momentum. This means that it ejects electrons from the Dirac sea and gives them positive momentum. However, as the Dirac sea has finite depth, there are only so many times $b\+_+(k)$ can be applied before it runs out of legal moves. In very excited states, the commutator \eqref{KM boson} will no longer retain its canonical form, allowing the raising operator to eventually also return zero.

Let $d(k)$ be the number of times $b\+_\pm(k)$ must be applied to make sure \emph{any} state is annihilated. This number is easiest to calculate at $k = 2k\_S - 1$. Here there is only one possible electron-positron pair with such a huge momentum difference, and hence $d(2k\_S - 1) = 2$.

On the other hand, the operator $b_+\+(1)$ increases the momentum of a single electron by one, provided this higher momentum mode is unoccupied. Thus, if $N^+ = 1$, it will take at least $2k\_S$ applications of $b_+\+(1)$ to make sure this electron has been pushed to the highest possible mode, $l = k\_S - 1$, and that the state was subsequently annihilated. If $N^+ = 2$, it will take at least $2(2k\_S - 1)$ moves to make sure the state is annihilated. For $N^+$ electrons this number is $N^+(2k\_S - N^+ +1)$, and it is maximized for $N^+ = k\_S$ to give $d(1) = k\_S(k\_S + 1)$. This is how many applications of $b_+\+(1)$ are needed to invert the Dirac sea and populate all modes at $l \geq 0$.

The monotonically decreasing function $d(k)$ gives an upper bound to the dimensionality of the scalar Hilbert space at momentum $k$. A marginally more precise counting reveals it to behave as
\bel{\label{m}
  d(k) \sim \left\{
           \begin{array}{ll}
             k\_S^2/k, & k \leq k\_S, \\
             2k\_S - k, & k \geq k\_S.
           \end{array}
         \right.
}

The upshot of this discussion is that, in the scalar theory dual to the free fermion, the \emph{size of the target space nontrivially depends on the momentum}. The dependence is roughly given by \eqref{m}. More precisely, there is a different dimension of the momentum $k$ Hilbert space in each superselection sector labeled by $N^\pm$. The result above shows the largest possible $d(k)$ for any $N^\pm$. For $k < k\_S$, the largest $d(k)$ is reached in the sector with $N^\pm = k\_S$, which is precisely the ``half-filling'' sector that contains the Dirac sea state.

A simple consequence of this momentum-dependence is that, for bosonization purposes, it does not actually matter what $K$ features in the lattice clock model \eqref{def H K}, as long as it satisfies
\bel{\label{inequality}
  K > \max_k \, d(k) = d(1) \sim k\_S^2.
}
If this condition is met in the microscopic theory, then the scalar theory possesses a large enough algebra to dualize to a fermion with ``string scale'' $k\_S$. In particular, this means that to get the desired bosonization, the scalar target space must be coarse-grained down to an effective size of $k\_S^2$ at most. If \eqref{inequality} is not satisfied, Abelian bosonization is inconsistent.

Note that the scalar side duality is defined on almost the same lattice as the fermionic side: there are $2k\_S - 2$ dynamical momentum modes and $M$ lattice points. Bosonization therefore implies that the target space has to be much bigger than the momentum space in order to discuss smoothness, since one has $K \gtrsim k\_S^2$. A hierarchy of this sort is in fact natural, as only in such a regime would the lattice see a continuous scalar field on each site.

The duality \eqref{duality basic} can also be \emph{twisted} by introducing background gauge fields in either theory. This procedure was discussed in great detail in \cite{Radicevic:2018okd} for exact lattice dualities. It lies beyond the present scope, however, as it requires a definition of smooth gauge fields. For now, suffice it to note that one of the twists will allow the bosonization of operators charged under $N^\pm$, i.e.~of individual fermions. This will be the formal way of deriving the ``Klein factors'' or solitons that enter the commonly cited formula $\Psi^\pm \sim \e^{\i \phi_\pm}$ \cite{Mandelstam:1975hb, Haldane:1979, Heidenreich:1980}. The other twist --- adding gauge fields to the fermion --- will result in a bosonization between the Schwinger model, or QED in $(1 + 1)$D, and a scalar field with a dynamical zero mode. This way one may expect to \emph{prove} the whole web of 2D dualities \cite{Karch:2019lnn}.

Another important aspect of the duality is the mapping of individual Hamiltonians. The original detailed mapping was derived by Haldane \cite{Haldane:1981zza}. It is clear that the interacting theory \eqref{def Lutt Model g h} will map to a quadratic theory of bosons, as long as only smooth operators are present in the Hamiltonian. Starting from the Luttinger model, the dual bosonic Hamiltonian will receive $k\_S/M$ corrections that should dramatically alter the Hamiltonian at momenta above $k\_S$. This will be reported elsewhere.

\section{Lessons learned about QFT}

Now that the dust has settled, it would be wise to revert back to more standard continuum notation and rephrase some key results in it. Most importantly, let the lattice spacing be
\bel{
  a \equiv \frac L M,
}
for some finite number $L$ that can be interpreted as the dimensionful length of the spatial circle. The string scale can also be associated to a dimensionful quantity,
\bel{
  \ell\_S \equiv \frac L {k\_S}.
}
Dimensionful position and momentum coordinates can be defined as
\bel{
  x\^c \equiv x a \in [0, L), \quad k\^c \equiv \frac{2\pi}L k \in \left[-\frac\pi a, \frac \pi a\right).
}
The smoothness scale is reached at momenta $k\^c\_S \equiv \pm 2\pi/\ell\_S$.

The continuum fermion fields are defined as
\bel{
  \Psi(x\^c) \equiv \frac1{\sqrt a} \Psi(x),
}
giving them an engineering dimension of $1/2$. This way the canonical anticommutation relation is
\bel{\label{anticomm wrong}
  \left\{\Psi^\alpha(x\^c)\+, \Psi^\beta(y\^c)\right\}  = \frac1a \delta_{x, y} \equiv \delta(x\^c - y\^c).
}
(But see eq.~\eqref{anticomm right}!) Note that matrix elements of continuum fermion fields are either 0 or $\frac1{\sqrt a}$.

The smoothness condition \eqref{Psi smoothness} can be written as
\bel{
  \Psi(x\^c + a) = \Psi(x\^c) + a \hat \del\^c \Psi(x\^c) + \ldots
}
where $\hat\del\^c$ is the formal derivative w.r.t.~$x\^c$. This way $\del\^c \Psi(x\^c) \equiv \frac1a[\Psi(x\^c + a) - \Psi(x\^c)]$ is restricted to have matrix elements that equal those of $\hat\del\^c \Psi$ and that hence diverge no faster than $1/\ell\_S\sqrt a$ in the continuum limit $a, \ell\_S \rar 0$. Without the smoothness condition, elements of $\del\^c\Psi(x\^c)$ could diverge as fast as $1/{a\sqrt a}$. The derivative corrections are thus controlled by the small parameter $a/\ell\_S$. This observation will be crucial when interpreting $\Psi(x\^c)$ as an operator-valued distribution and integrating it against smooth functions.

Looking back, the main motif in this paper was the following question: what does a product of continuum fields mean? The discussion of subsection \ref{subsec ope} shows that there are really two answers to this question. One option is to naively multiply the continuum fields, defining e.g.~the product of a fermion and its conjugate to be
\bel{\label{prod IR}
  \Psi^\alpha(x\^c)\+  \Psi^\beta(y\^c).
}
This operator has no information about the lattice dynamics at $|x\^c - y\^c| \ll \ell\_S$, as shown by eq.~\eqref{Psi Psi aux}. In fact, most such products will be trivial, with the exception of those that get to detect an RG invariant such as the chiral anomaly, as exemplified by eq.~\eqref{KM pos space}.

The other option is to acknowledge that all continuum fields come from lattice fields via the smoothing procedure. Thus, one can try to define an uplift map
\bel{
  \Psi^\alpha(x\^c) \mapsto \Psi_x^\alpha,
}
and then define the product of fermion fields as the smoothing of the product of their uplifts,
\bel{\label{prod UV}
  (\Psi^\alpha)\+ \Psi^\beta(x\^c, y\^c).
}
This construction does know about correlations below the ``string scale,''  as shown in eq.~\eqref{OPE Dirac}. In turn, these correlations can be understood to \emph{define} the continuum theory by supplying it with this product structure. The problem, however, is that the uplift map is not unique. In a general CFT, for instance, it may not be obvious how to uplift a given primary operator.

Having said all this, which product do field theorists use when multiplying two continuum fields? The answer is that \emph{both} are used, often without specifying which choice was made. For instance, the anticommutation relation \eqref{anticomm wrong} was intentionally written wrong; it should actually say
\bel{\label{anticomm right}
  \left\{(\Psi^\alpha)\+, \Psi^\beta \right\}(x\^c, y\^c)  = \frac1a \delta_{x, y} \equiv \delta(x\^c - y\^c).
}
Have you noticed? This illustrates how easy it is to slip and use inconsistent notation by sheer force of habit. On the other hand, the commutator of currents \eqref{KM all momenta} that gives rise to the Kac-Moody structure was not written incorrectly: here it is really necessary to compute the product of smoothed currents, not the smoothing of the product of currents. There does not appear to exist a simple way to distinguish between the two uses in the literature. Hopefully the formulation of the two possible options presented in this paper will make it easier to consistently address ambiguities in operator products, without having to invoke ambiguous procedures such as normal-ordering.

Knowing how to multiply continuum fields is equivalent to knowing when to insert an OPE, defined in a general setting by eq.~\eqref{OPE}. This paper has shown that OPEs can be defined in arbitrary theories, and that this definition reproduces the CFT forms for $\Psi\+ \times \Psi$ and $J \times J$ OPEs directly from the lattice, in eqs.~\eqref{OPE Dirac} and \eqref{OPE JJ}. A few more comments are in order here. The constant prefactors in these formulae can be changed by choosing a particular scaling of the continuum operators, so they have not been compared to the literature. More mystifying may be the wildly oscillating term $1 - (-1)^{x - y}$ that stems from the sharp edge of the Dirac sea. If all continuum fields are interpreted as distributions to be integrated against smooth functions, such a term would be averaged out and replaced with a constant. Thus, for instance, the $\Psi\+ \times \Psi$ OPE in the continuum can be written as
\bel{
  (\Psi^\alpha_x)\+ \times \Psi^\beta_y = \frac{\i \delta^{\alpha\beta}}{2\pi(x\^c - y\^c)} + O\left( \frac{|x\^c - y\^c|}{\ell\_S}\right).
}
Textbooks typically write that the OPE is valid at $x\^c \rar y\^c$, and this lattice computation makes this limit precise. This OPE is the term that needs to be added to the product of two smooth operators in order to include their UV correlations, i.e.~their correlations when the insertion points are less than $\ell\_S$ away.

Going full circle, knowing how to multiply continuum operators is also equivalent to knowing the precise lattice-continuum correspondence described at the beginning. Using this knowledge, this paper has derived Abelian bosonization using operator techniques, at the same level of rigor one can derive Kramers-Wannier duality. (To wit: completely rigorously.) It is remarkable that we may now be in the position to rigorously prove IR dualities.

A large number of questions remain open. Perhaps the most glaring omission in this paper is the lack of a discussion of smooth bosonic operators. This subject is more involved because bosons have another large parameter, the target space size $K$, and its synergy with the ``string scale'' is still mysterious. (An early attempt to tame the target space using RG techniques is in ref.~\cite{Radicevic:2016kpf}.) A particularly intriguing question is whether ``gauging'' refers to including gauge fields before or after smoothing. The latter option may be an elegant way to avoid the geometric anomaly \cite{Radicevic:2018zsg}.

There do exist other appealing research directions. Generalizing this analysis to multiple fermionic flavors \cite{Witten:1983ar} and connecting it to higher-dimensional bosonization \cite{Chen:2017fvr} appears within reach. It would also be interesting to identify and compute the energy-momentum tensor OPEs on the lattice. A different direction would be to understand the path integral description of the smoothness condition, and to ultimately connect this work with the ideas in algebraic QFT. For instance, it would be nice to prove that the sectors appearing in the algebra of smooth operators, or at least the Dirac sector, correspond to type III algebras in the thermodynamic limit.

\section*{Acknowledgments}

I am grateful to Bruno Le Floch, Nabil Iqbal, Sunil Mukhi, Erich Poppitz, Shinsei Ryu, and Steve Shenker for useful discussions. The majority of this work was done at the Perimeter Institute for Theoretical Physics, which is supported by the Government of Canada through Industry Canada and by the Province of Ontario through the Ministry of Economic Development \& Innovation. Parts of this work were carried out (and/or presented) at the Aspen Center for Physics (supported by National Science Foundation grant PHY-1607611), at the Yukawa Institute for Theoretical Physics at Kyoto University (during the \emph{Quantum Information and String Theory} workshop supported by the Simons Foundation), and at the International Center for Theoretical Physics in Trieste (during the \emph{New Pathways in Explorations of Quantum Field Theory and Quantum Gravity Beyond Supersymmetry} workshop).  The final stages of this work were completed with support from the Simons Foundation through \emph{It from Qubit: Simons Collaboration on Quantum Fields, Gravity, and Information}, and from the Department of Energy Office of High-Energy Physics through Award DE-SC0009987.

\bibliographystyle{ssg}
\bibliography{ABRefs}

\end{document}